\documentclass[manuscript,screen,acmsmall]{acmart}

\def\markup{0}

\if\markup1
\usepackage[normalem]{ulem}
  \definecolor{myblue}{rgb}{0,0,0.75}
  \newcommand{\rv}[1]{{\leavevmode\color{myblue}#1}}
  \newcommand{\st}[1]{{\sout{#1}}}
\else
  \newcommand{\rv}[1]{#1}
\newcommand{\st}[1]{}
\newcommand{\sout}[1]{}
\fi

\usepackage{array}
  
\AtBeginDocument{%
  \providecommand\BibTeX{{%
    \normalfont B\kern-0.5em{\scshape i\kern-0.25em b}\kern-0.8em\TeX}}}

\setcopyright{acmcopyright}
\copyrightyear{2018}
\acmYear{2018}
\acmDOI{XXXXXXX.XXXXXXX}

\acmConference[Conference acronym 'XX]{Make sure to enter the correct conference title from your rights confirmation email}{June 03--05, 2018}{Woodstock, NY}
\acmPrice{15.00}
\acmISBN{978-1-4503-XXXX-X/18/06}

\begin{document}

\title[AI as a Bridge Across Ages]{AI as a Bridge Across Ages: Exploring The Opportunities of Artificial Intelligence in Supporting Inter-Generational Communication in Virtual Reality}


\author{Qiuxin Du}
\orcid{0009-0000-5991-9482}
\authornote{Both authors contributed equally to this research.}
\affiliation{%
  \institution{Beijing Institute of Technology}
  \city{Beijing}
  \country{China}
}
\email{qiuxindu@bit.edu.cn}

\author{Xiaoying Wei}
\orcid{0000-0003-3837-2638}
\authornote{Both authors contributed equally to this research.}
\affiliation{%
 \institution{The Hong Kong University of Science and Technology}
 \city{Hong Kong SAR}
 \country{China}
 }
\email{xweias@connect.ust.hk}

\author{Jiawei Li}
\orcid{0009-0000-6593-2958}
\affiliation{%
  \institution{The Hong Kong University of Science and Technology (Guangzhou)}
  \city{Guangzhou}
  \country{China}
}
\email{crane787213823@outlook.com}

\author{Emily Kuang}
\orcid{0000-0003-4635-0703}
\affiliation{%
  \institution{ Rochester Institute of Technology}
  \city{Rochester}
  \country{USA}
}
\email{ek8093@rit.edu}

\author{Jie Hao}
\orcid{0000-0003-1730-4847}
\affiliation{%
  \institution{Beijing Institute of Technology}
  \city{Beijing}
  \country{China}
}
\email{haoj@bit.edu.cn}

\author{Dongdong Weng}
\orcid{0000-0003-2352-0896}
\affiliation{%
  \institution{Beijing Institute of Technology}
  \city{Beijing}
  \country{China}
  }
\email{crgj@bit.edu.cn}

\author{Mingming Fan}
\orcid{0000-0002-0356-4712}
\authornote{Corresponding Author}
\affiliation{%
  \institution{The Hong Kong University of Science and Technology (Guangzhou)}
  \city{Guangzhou}
  \country{China}
}
\affiliation{%
 \institution{The Hong Kong University of Science and Technology}
 \city{Hong Kong SAR}
 \country{China}
 }
\email{mingmingfan@ust.hk}

\renewcommand{\shortauthors}{Paper id: 7553}

\begin{abstract}
 Inter-generational communication is essential for bridging generational gaps and fostering mutual understanding. However, maintaining it is complex due to cultural, communicative, and geographical differences. Recent research indicated that while Virtual Reality (VR) creates a relaxed atmosphere and promotes companionship, it inadequately addresses the complexities of inter-generational dialogue, including variations in values and relational dynamics. To address this gap, we explored the opportunities of Artificial Intelligence (AI) in supporting inter-generational communication in VR. We developed three technology probes (e.g., Content Generator, Communication Facilitator, and Info Assistant) in VR and employed them in a probe-based participatory design study with twelve inter-generational pairs. Our results show that AI-powered VR facilitates inter-generational communication by enhancing mutual understanding, fostering conversation fluency, and promoting active participation. We also introduce several challenges when using AI-powered VR in supporting inter-generational communication and derive design implications for future VR platforms, aiming to improve inter-generational communication.
\end{abstract}



\maketitle

\section{Introduction}

 Inter-generational communication refers to the exchange of experiences and perspectives across generations, fostering a deeper mutual understanding and bridging the gap between different generations within a family~\cite{williams2013intergenerational, kemp2005dimensions, taylor2005distance}. Inter-generational communication plays a crucial role in nurturing emotional growth, enhancing self-esteem, and cementing familial bonds for younger members~\cite{brussoni1998grandparental, Tomlin1998handbook, williams1997young}, while simultaneously providing older generations with essential social engagement and a cherished role in the family structure~\cite{lindley2009desiring, cornejo2013enriching, seifert2020digital, nicholson2012review, cotterell2018preventing}.  
 However, modern society's rapid technological and cultural shifts pose considerable challenges to maintaining effective inter-generational communication. The disparity in cultural backgrounds, communication styles, and subtle emotional expressions, coupled with geographical distances, often hamper the flow and depth of these interactions~\cite{stafford2004maintaining, vutborg2010family}. \sout{As such, the need to facilitate more active and effective inter-generational dialogue has become a pressing concern, especially within the HCI and CSCW communities.}

 Recent studies have seen the emergence of virtual reality (VR) as a promising medium to support inter-generational communication. This platform offers unique affordances, such as customizable avatars, immersive virtual environments, and a sense of co-presence, which have shown promise in fostering more relaxed, equal, and emotionally fulfilling inter-generational interactions~\cite{baker2021school, li2019measuring, wei2023bridging, hoeg2021buddy}. Studies also showed that communication via VR could improve the overall quality of life for older people while decreasing the caregiver burden of their younger generations~\cite{afifi2022using, hoeg2021buddy}. While previous research has explored various aspects of VR, they often focused on predefined features and specific usage contexts~\cite{afifi2022using, hoeg2021buddy, wei2023bridging, li2023exploring}, not fully addressing the complexities of inter-generational dialogue, such as differing values, generational hierarchy, evolving emotional states, or relational dynamics (e.g., conflict management)~\cite{devito2007interpersonal, vutborg2010family}. 
 
 Previous research has shown artificial intelligence's (AI) potential to bridge these gaps by understanding and responding to users' dialogue content and emotions in real-time~\cite{zheng2022ux, benke2020chatbot}. This includes technologies like large language models (LLMs), which can comprehend and generate human-like text, and AI-generated content (AIGC), which involves creating various types of media using AI. \sout{Several studies have} Researchers leveraged these technologies to support inter-generational communication in non-immersive environments~\cite{kang2021momentmeld, kang2023ai, zhang2022storybuddy, Isbister2000helper}, revealing AI's ability to manage complex communication dynamics and promote reminiscence~\cite{zheng2022ux, kang2021momentmeld, kang2023ai, zhang2022storybuddy}. However, there is limited understanding of how AI can be used in VR to support inter-generational communication. VR offers a flexible platform for AI to present diverse outputs (e.g., text, audio, images, 3D models) through interactive avatars and immersive environments, while allowing users to physically interact with AI-generated objects~\cite{gandedkar2021role, manolakis2022virtual, herron2016augmented, bussell2023generative, schulenberg2023towards}, potentially transforming inter-generational interactions. To understand how inter-generational pairs would use AI in VR to facilitate communication, we propose our first research question, \textbf{RQ1: How could AI-powered VR be leveraged to support inter-generational communication}? 
 
 Furthermore, previous works have shown several challenges in using AI for interpersonal communication. For instance, AI interventions could disrupt the smooth flow of group discussions~\cite{maier2022comparing, li2023improving, duan2021bridging}. Unlike formal communication, inter-generational communication within families tends to be more casual and focused on personal topics, often reflecting familial backgrounds~\cite{lin2002conversation, heshmat2021family}. This informal and personal communication style could present new challenges for AI in understanding and appropriately intervening in inter-generational exchanges. This leads us to explore our second research question: \textbf{RQ2: What are the challenges of using AI to support VR-based inter-generational communication?} 

 Building on previous research that demonstrated the effectiveness of technology probe-based participatory design in eliciting user feedback and identifying opportunities in emerging technologies~\cite{pradhan2020understanding, li2023exploring, ARforSmartphone}, \sout{such as older adults' perceptions of the Internet of Things technologies and AR for inter-generational storytelling} we adopted a similar methodology. Using three types of AI probes (e.g., Content Generator, Communication Facilitator, and Info Assistant), we conducted a probe-based participatory design study with twelve inter-generational pairs \sout{(e.g., grandparent-grandchild, parent-child, uncle-niece)} to uncover the effectiveness and user expectations of AI-powered VR in supporting communication. Participants first tried out these AI probes to inspire them to think about the possible ways in which AI could enhance their communication in VR. They then participated in a co-design workshop to brainstorm and design desired AI features to assist their communication.
 
 Our findings illustrate the multifaceted advantages of AI-powered VR in supporting inter-generational communication. Drawing from the outcomes of our observation and participatory design workshop, participants leveraged AI-powered VR to 1) foster mutual understanding by transforming abstract conversation content into concrete forms for better idea exchange, along with offering various ways to interact with the generated content to deepen communication; 2) enhance conversation fluency by personalizing AI representations for better acceptance, and providing interventions to prevent awkward silences, maintain balanced dialogue, and manage potential frictions; 3) promote their active participation by modifying users' avatars or virtual environments based on their emotions and contextual contents. Our results also reveal the challenges of using AI-powered VR to support inter-generational communication. Based on the results, we proposed design implications for future designers to better leverage AI to support VR-based communication.

 In sum, we make the following contributions:

\begin{itemize}
    \item Our study advances the understanding of AI-powered VR in promoting inter-generational communication. We demonstrate AI's effectiveness in enhancing mutual understanding, improving conversational fluency, and encouraging active participation, highlighting its role in bridging generational gaps in VR interactions.
    \item Drawing from participatory design outcomes, we outline participants' insights into customizing AI features to address specific communication needs in inter-generational interactions, as well as provide design implications for future researchers and developers to create more practical and meaningful AI-based VR platforms.
\end{itemize}

\section{Related Work}

\subsection{Benefits and challenges of inter-generational communication within a family}
\label{sec:challenges}
 Inter-generational communication involves exchanging information, beliefs, and feelings between individuals from different generations. Inter-generational communication within families (i.e., connection across generations like grandparents, parents, and children) is particularly beneficial~\cite{williams2013intergenerational, kemp2005dimensions, taylor2005distance}. 
 On the one hand, \sout{engaging in close inter-generational communication} \rv{it} supports the growth and emotional development of younger generations, enhancing their self-esteem and deepening familial bonds~\cite{brussoni1998grandparental, Tomlin1998handbook, williams1997young}. On the other hand, it \sout{is crucial for older generations,} provide \rv{older generations} with valuable social engagement and reducing feelings of isolation~\cite{lindley2009desiring, cornejo2013enriching, seifert2020digital, nicholson2012review}. This helps maintain an active and cherished role within the family dynamic. \sout{Thus, close inter-generational communication not only bridges the generational gap but also enriches the family as a whole, fostering a deeper understanding and appreciation of each generation's unique contributions~\cite{seponski2009caring, kornhaber1981grandparents}.}
 
 However, maintaining successful inter-generational communication is challenging in today's society, which is characterized by rapid technological and cultural shifts.
 First, while people have a strong need and desire to stay connected with their families to share recent life events, two generations find it \textit{difficult to achieve mutual understanding} due to disparate cultural backgrounds, including distinct values, beliefs, and communication styles~\cite{mynatt2001digital, tee2009exploring}. This often leads to misinterpretations and conflicts in their exchanges\sout{, as well as reducing active engagement during conversation}~\cite{romero2007connecting, neustaedter2006interpersonal}.
 Second, the generational hierarchy often results in the younger generation demonstrating \textit{over-accommodations} towards the elderly~\cite{kemper2000accommodations}, \sout{Over-accommodation refers to the excessive or inappropriate adjustment of one's communication style, language, or behavior to cater to another person's perceived needs,} \rv{such as listening passively without interrupting.} Although these actions are well-intentioned and display respect for the older generation, they frequently lead to miscommunication, a lack of genuine connection, and frustration among young people~\cite{williams1996intergenerational, giles2003intergenerational}.
 Third, \textit{the subtlety of emotional expression} between two generations hinders successful inter-generational communication, especially in countries with Eastern cultures~\cite{kincaid2013communication, StudyMassCom2023}. \sout{For example, older adults may express concern or affection through actions or indirect language, such as providing unsolicited advice or preparing meals, rather than direct verbal expressions like 'I care about you.' Younger generations, accustomed to more direct communication, might misinterpret these gestures as overbearing or unnecessary interference, rather than recognizing them as expressions of love and concern.}
 
 Additionally, geographical distance between family members often impedes intimate inter-generational communication~\cite{forghani2014routines, heshmat2021family}. While audio-visual tools like audio or video calls enable remote connections, they do not always provide enough emotional support to give a feeling of closeness and intimacy.
 Therefore, further study on how to facilitate more active and effective inter-generational communication is warranted by HCI researchers.

\subsection{VR for supporting inter-generational communication}
\label{sec:VRchallenge}
Previous works in the HCI field have leveraged VR to support inter-generational communication~\cite{li2023exploring, wei2023bridging, afifi2022using, hoeg2021buddy}. These studies demonstrated three affordances of VR that can support communication between two generations. First, VR allows users to represent themselves via customizable avatars. Such representation affects their perceptions of each other and blurs the generation gaps, thereby fostering a more casual and equal communication experience than in reality~\cite{wei2023bridging}. Additionally, the study indicated that users could engage in various activities in a virtual environment, which stimulates more topics for them to keep their conversations flowing~\cite{wei2023bridging, wang2022construction, hoeg2021buddy, afifi2022using}. Finally, VR's immersive environment offers the sense of co-presence for two generations to achieve a degree of companionship~\cite{abdullah2021videoconference, li2019measuring}, which is important for maintaining close emotional connections.
Researchers also showed that using VR as a long-term communication tool improved the overall quality of life for older people while decreasing the caregiver burden of younger generations~\cite{afifi2022using} \sout{who live far away}. 

While previous studies have demonstrated the potential of VR in facilitating inter-generational communication, their emphasis has predominantly been on exploring predefined features to meet user needs within specific communicative contexts. For instance, Wang et al. \cite{wang2022construction} and Hoeg et al. \cite{hoeg2021buddy} explored the design of collaborative activities (e.g., tandem biking, co-learning) in VR for two generations. Wei et al. investigated the design of avatar appearances in VR to facilitate their communication~\cite{wei2023bridging}. However, these predefined designs fall short of addressing the myriad complexities inherent in inter-generational dialogue, such as navigating differing knowledge, over-accommodation, and the evolving nature of users' conversational content and emotional states~\cite{mynatt2001digital, tee2009exploring, williams1996intergenerational, giles2003intergenerational}. Therefore, this work explores how to design an interactive VR environment that effectively addresses the challenges present during inter-generational communication in VR.

\subsection{Integrating AI into inter-generational communication within VR environments}
\label{sec:DC}
Previous research has demonstrated the potential of AI to moderate inter-generational communication by understanding \sout{users'} dialogue's content and emotions in real-time, and intervening accordingly~\cite{zheng2022ux, benke2020chatbot}. Studies have leveraged AI to support inter-generational communication in non-immersive environments, \sout{These studies used AI} including providing context-related photos to arouse reminiscence and foster conversation~\cite{kang2021momentmeld, kang2023ai}, generating story-related questions to enhance learning outcomes~\cite{zhang2022storybuddy, entenberg2023ai}, and answering users' questions to improve the collaboration efficacy~\cite{lee2022interactive}. 

While AI shows potential in enhancing inter-generational communication, its integration into immersive VR environments remains under-explored. The integration of AI and VR has the potential to facilitate inter-generational communication for two main reasons. First, AI could enable VR to dynamically adjust its environment and settings by monitoring users' dialogue in real-time, potentially addressing the limitations associated with predefined designs in previous VR research~\cite{duan2021bridging, li2023improving, luck2000applying, zheng2022ux}. Second, unlike non-immersive tools that offer limited interaction (e.g., text-based or speech-based agents~\cite{zheng2022ux}), immersive VR environments present a unique opportunity to leverage various advanced AI outputs to enhance communication, including interactive and responsive virtual avatars, dynamic 3D visualizations, and adaptable virtual spaces~\cite{gandedkar2021role, manolakis2022virtual, herron2016augmented, bussell2023generative, schulenberg2023towards}. This combination could potentially address the inter-generational communication challenges presented in Section~\ref{sec:challenges}, \rv{such as providing various AI-generated content (e.g., text, images, objects) to exchange ideas for achieving mutual understanding.} 
\sout{For instance, to achieve \textit{mutual understanding}, VR could provide users with various AI-generated content (e.g., text, images, objects) and allow users to interact with them, helping them better exchange ideas~\cite{al2001bridging}. To reduce \textit{over-accommodations} in conversations, VR could provide appropriate forms of AI interventions when detecting the conversations becoming stagnant or imbalanced. To enhance the \textit{emotional expression}, VR could visualize and exaggerate users' emotional expressions using avatars or the surrounding environment when AI detects emotional changes~\cite{lee2022understanding}. }

Despite these potential benefits, little is known about how inter-generational pairs perceive and leverage AI in VR to support communication, as well as the challenges and considerations in designing AI features for better supporting inter-generational communication. Understanding these aspects could help future researchers and developers create more practical and meaningful AI-based VR platforms. To address these gaps, we employed a probe-based participatory design approach to explore the effectiveness and user perceptions of AI in enhancing inter-generational communication in VR.

\subsection{Technology probe methodology}

A technology probe study is an innovative research methodology primarily used in the fields of HCI and CSCW~\cite{huang2014wearable, hutchinson2003technology, li2023exploring}. This method typically offers fundamental functionalities of specific emerging technology, enabling users to explore and understand their interactions with these innovations~\cite{hutchinson2003technology}. The primary purpose of a technology probe study is to inspire and provoke responses from users, shedding light on user needs, desires, and technological possibilities, rather than to test a specific hypothesis or to refine a product for real-life use. This is particularly valuable in the early stages of technology development, where understanding the user experience is crucial~\cite{pradhan2020understanding, li2023exploring}. 

Prior work has shown the feasibility of the technology probe study in identifying opportunities of emerging technology in inter-generational communication, such as AR~\cite{li2023exploring} and VR~\cite{wei2023bridging}. By observing how participants use and adapt to the technology, these studies identify unforeseen opportunities and challenges, leading to more user-centered design decisions. Therefore, in this work, we adopted the technology probe study method to explore AI's possibilities for facilitating inter-generational communication in immersive environments and answering our RQs.

\section{Method}

To answer our RQ, we designed and implemented AI probes, and then conducted a probe-based participatory design workshop to elicit participants' perceptions and envisions \sout{of AI designs}. For simplicity, in the rest of this paper, the older and younger generation are abbreviated as OG and YG respectively (e.g., P1-YG means the younger generation in pair 1).

\subsection{Probe Design and Implementation}
\begin{figure}[]
\centering
\includegraphics[width=0.9\linewidth]{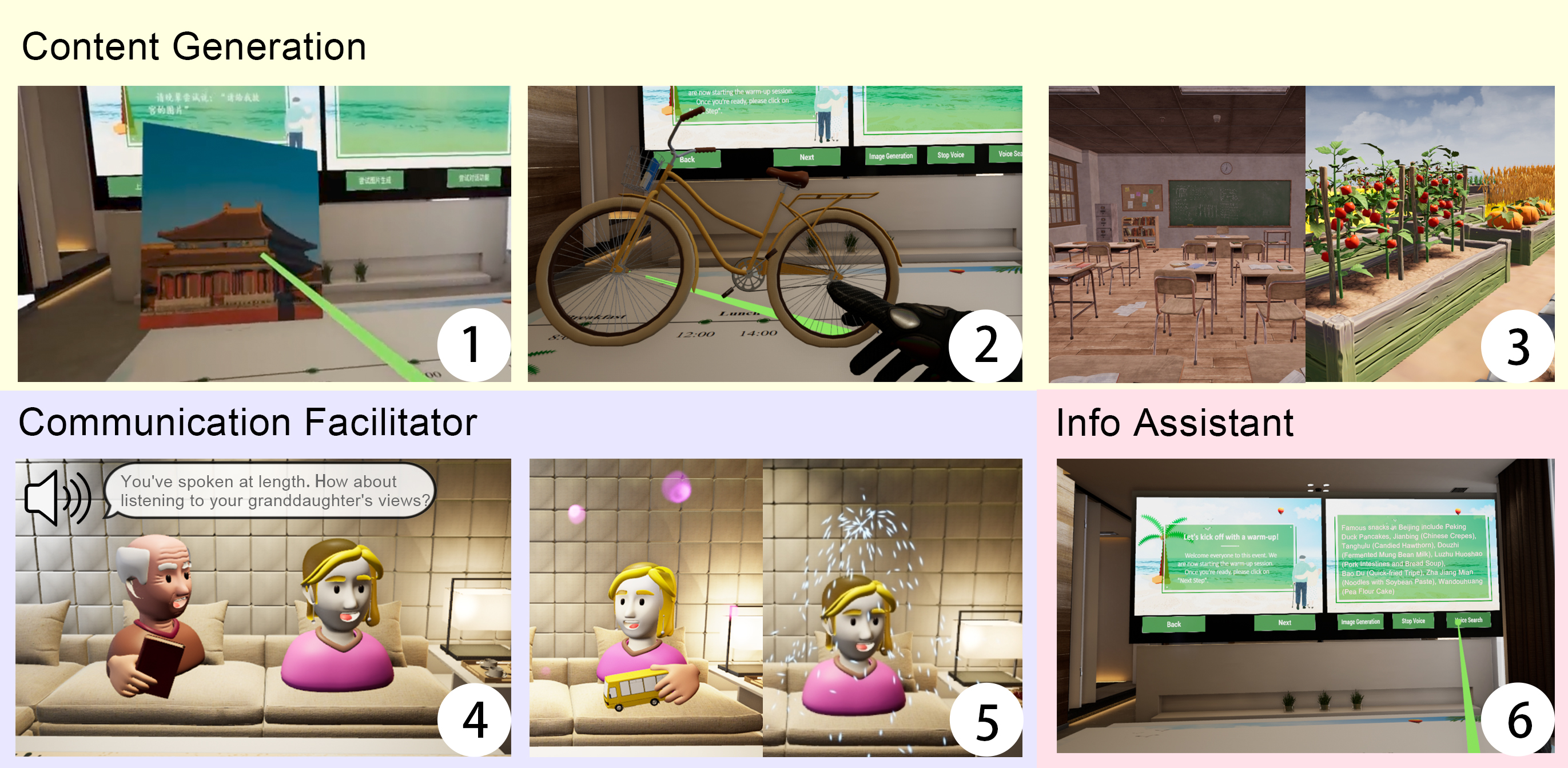}
\vspace{-0.2cm}
\caption{Illustrations of our AI probes. (1) Image generation: it generates images based on verbal descriptions provided by users. (2) 3D Object generation: it creates virtual 3D models according to users' verbal descriptions. (3) Scene generation: the virtual environment automatically switches based on the content of the conversation. (4) Chat facilitator: it intervenes verbally to balance conversations when one party dominates, and offers potential conversation topics through voice prompts during awkward silences. (5) Emotion visualization: analyzing the emotional tone of chat content, it displays bubble effects for positive emotions (left) and rain effects for negative emotions (right). (6) Info assistant: users press the button to input voice-based queries and receive spoken and text answers in response to their inquiries.} 
\Description{This figure illustrates the AI probes used in our study. The first image is about 'Image generation', showing our system generates a photo of Forbidden City based on verbal descriptions. The second image is about '3D Object generation', showing that users create a 3D bike model according to their verbal descriptions. The third image is 'Scene generation', showing that the virtual environment in VR automatically switches to an Old classroom or Farmland based on the content of the conversation. The fourth is 'Chat facilitator', demonstrating that our system intervenes verbally to balance conversations by saying 'How about listening to your granddaughter's opinions?' The fifth image is about 'Emotion visualization' showing that our system displays bubble effects for positive emotions (left) and rain effects for negative emotions (right). The sixth image is about the 'Info Assistant', showing that users press the button to input voice-based queries and receive spoken and text answers in response to their inquiries.} 
\label{fig:probe}
\end{figure}

We \sout{designed and }developed three types of AI probes, including \textbf{Content Generation} (i.e., image generation, 3D object generation, and scene generation), \textbf{Communication Facilitator} (i.e., chat facilitator and emotional visualization), and \textbf{Info Assistant}, as shown in \autoref{fig:probe}. The probes include three types of AI outputs: text/voice, images, and 3D objects/environments, and can be activated in two ways: passively and proactively. We provide these probes to offer participants fundamental and comprehensive AI functionalities to experience in VR~\cite{hutchinson2003technology}, which allows us to understand how inter-generational pairs interact with AI while communicating in VR, what challenges they may encounter, and how they envision future AI features to be designed in VR to better support their communication. In the following section, we introduce the design considerations and implementation details of each AI probe.

\subsubsection{Content Generation} As discussed in Section \ref{sec:DC}, VR's capability to showcase various AI outputs enables users to visualize their conversation content and interact with it, \sout{This capability potentially} potentially enhancing mutual understanding and bridging knowledge gaps across generations~\cite{al2001bridging}. To explore how inter-generational pairs leverage Content Generation in VR to facilitate communication \sout{and identify the challenges they encounter}, we present three fundamental forms of AI outputs \sout{as technical probes }to stimulate users' imagination and insight.

\textbf{Image generation.} Images can effectively convey information such as emotions, concepts, and events, playing a crucial role in facilitating communication in daily life~\cite{ballenger2014photography, kang2021momentmeld}. A previous study demonstrated that sharing and discussing images in a VR environment provided higher feelings of co-presence and increased the quality of interaction compared to screen-based communication (e.g., Skype)~\cite{li2019measuring}. 
In this work, we propose an Image generation function in VR. \sout{to explore how users leverage this feature to support inter-generational communication.} This feature is activated when users click a button and provide verbal descriptions of an image. The system then generates the image based on these voice prompts and displays it in front of two users. Users can manipulate these images by grabbing, moving, zooming, or rotating with their hands, \sout{which are unique interactions in VR} as illustrated in the first image of~\autoref{fig:probe}. If the generated images do not meet the users' expectations, they can be discarded by dragging them into a virtual trash bin. For implementation, we used Azure AI Speech for voice transcription~\footnote{https://azure.microsoft.com/en-us/products/ai-services/ai-speech} and DALL-E2 for image generation~\footnote{https://openai.com/dall-e-2}.

\textbf{3D object generation.} 3D objects are particularly useful for explaining specific objects or concepts \sout{in daily conversation}~\cite{di20153d}. Considering the rendering time and computing costs of 3D object generation from text, we opted to provide users with a selection of pre-generated 3D objects relevant to the tasks, such as various vehicles (bus, taxi, etc.) for the first task and assorted stationery items (vintage chair, ruler, pencil case, printer, etc.) for the second, as shown in \autoref{fig:3dobjects}. For implementation, we used Point-E to generate 3D models from text~\cite{nichol2022point}. Additionally, we implemented a keyword detection that automatically activates this feature whenever users mention specific terms from our curated list. Upon detection, these 3D objects materialize in front of the users. Users can interact with these 3D objects by grabbing, moving, zooming, or rotating them.

\textbf{Scene generation.} A content-specific environment creates an engaging and active communication atmosphere for inter-generational users~\cite{wei2023bridging}. Similar to 3D object generation, we provided pre-created virtual scenes pertinent to the tasks, such as a living room or a farm for the first task and both a new and an old classroom for the second, as shown in \autoref{fig:scene} in the Appendix. These scenes are activated through keyword recognition. The scenes were developed using Set-the-Scene~\cite{cohen2023set}.

\subsubsection{Communication Facilitator} 
Integrating an AI coordinator to monitor dialogue and offer timely interventions \sout{has the potential to} enhance communication efficiency~\cite{zheng2022ux, duan2021bridging, li2023improving}. To explore the possibilities of VR to support AI coordinators in inter-generational contexts, we developed the chat facilitator (voice-based intervention) and emotional visualization (visual-based intervention).

\textbf{Chat Facilitator.} As discussed in Section~\ref{sec:DC}, \textit{over-accommodation} and \textit{the lack of conversation topic} often hinder effective inter-generational communication. Therefore, we provided verbal interventions. We detected \textit{imbalances} (i.e., one party speaks continuously for over one minute while the other remains silent) and \textit{silences} (i.e., no one speaks for more than thirty seconds) by recording the length of silence in real-time, \sout{of both parties during the conversation} and utilized prompt engineering \cite{kojima2023large} with ERNIE Bot~\footnote{https://yiyan.baidu.com/} to provide appropriate voice suggestions.\sout{to mitigate these issues} For example, if one party dominates the conversation, the Chat Facilitator might suggest, \textit{``How about listening to the other's opinions?''}

\textbf{Emotional Visualization.} We also provide visual interventions in VR\sout{in conversations} to help inter-generational users in \textit{expressing their emotions more effectively}. We employed keyword monitoring to detect positive (e.g., `happy', `excited') and negative emotions (e.g., `sad', `tired'). Subsequently, we incorporated specific visual effects that corresponded with these emotions. For example, the rain effect was chosen to express their negative emotions, as rain is usually associated with sadness. Conversely, bubble effects were used for positive emotions, reflecting their joyful connotations, as shown in the fifth image in~\autoref{fig:probe}.

\subsubsection{Info Assistant}
Prior studies showed that people often queried information with AI to support interpersonal communication, helping them communicate more confidently and accurately~\cite{fu2023text, duan2021bridging}. 

\textbf{Info Assistant.} This feature serves two main purposes: (1) to help users easily query information within the VR environment, supporting their daily conversations and experimental tasks; and (2) to allow users to experience voice interactions with AI in VR, inspiring them to envision how AI voice assistants can be better designed in VR, including interaction methods and AI representation.
Users can click a button and verbally query the AI, with responses displayed as text on the screen and simultaneously vocalized using Azure AI Speech, as depicted in the sixth image of~\autoref{fig:probe}. Participants can halt the audio playback at any time by activating the ``Stop Speech'' button. The search engine functionality is powered by ChatGPT-3.5~\footnote{https://chat.openai.com/}.

\subsection{System Design and Implementation}

Our goal was to integrate all the AI probes into a user-friendly VR application. We developed the system using Unreal Engine 5.1.1~\footnote{https://www.unrealengine.com/en-US/}. Our system supports remote VR communication through UDP (User Datagram Protocol) for real-time interaction and synchronization of virtual items and scenes across both users' interfaces. Additionally, we equipped the system with a camera to document participants' behaviors and interactions within the VR environment.

For the VR environment, we selected a living room as the task setting, complete with a sofa, coffee table, and typical household decorations, as illustrated in \autoref{fig:scene} in the Appendix. This cozy environment was chosen to provide users with a familiar and relaxing atmosphere to facilitate their communication~\cite{wei2022communication}. We set up two virtual screens to display different information. The left screen displays text instructions to 1) help users better learn the AI probe during the training phase and 2) guide them through each task during the study session. The right screen offers an interactive interface for 1) invoking Content Generation and Info Assistance functions, and 2) providing a display area for the results of Info Assistance, making it easily viewable for users.

For the avatars, we provided users with four types of avatars to embody in VR: an elderly male, an elderly female, a young male, and a young female, as shown in \autoref{fig:avatar} in the Appendix. Users were required to select one avatar that best matched their real-life identity to mitigate novelty effects~\cite{rodrigues2022gamification, miguel2024evaluation}. We did not offer an avatar customization function to mitigate distractions related to avatar appearance and focus on interacting with AI in VR. Additionally, to foster an active and relaxing atmosphere that encourages user interaction, we offered abstract and cartoon-style avatars instead of realistic ones~\cite{wei2023bridging}.

\subsection{Participants}

We advertised our recruitment information through snowball sampling, social media, and community centers. Before the study, participants completed an online recruitment survey. Based on their responses, we invited participants with diverse backgrounds (e.g., education level, occupation) to participate in the study. We ensured all participants met the inclusion criteria, which were no major medical illnesses, no history of mental or neurological abnormalities, and no sensory impairments. The university's Institutional Review Board (IRB) approved this study for research ethics.

We recruited 12 pairs of participants (24 in total) from two different regions in China (four pairs were from the metropolitan center, while eight were from regional locations). The YG participants were aged 15 to 30 ($M = 20.6, SD = 5.2$), and the OG participants were aged 55 to 78 ($M = 63.2, SD = 6.0$). The OG and YG participants in each pair had lived separately for more than one year and did not currently live together. Each pair had diverse relationships (6 pairs were grandparent-grandchild, 2 pairs were parent-child, and 4 pairs were other forms of kinship, such as uncle-niece). Most participants (OG: n=9, YG: n=12) had experience using AI-based products, including voice assistants (e.g., Siri or Tmall Genie), and generative AI applications (e.g., ChatGPT or ERNIE Bot). Most participants (OG: n=11, YG: n=9) had no prior experience with VR. Detailed participant background information is shown in \autoref{tab:participants}. 

\begin{table}[h]
 \caption{Demographic information for all participants}
  \label{tab:participants}
  \scalebox{0.85}{
  \begin{tabular}{ccccc}
    \toprule 
     Pair ID & Relationship & Age (OG-YG) & VR experience (OG-YG) & AI experience (OG-YG) \\
    \midrule 
    1 & GP-GC & 61-15 & No-No & No-Yes \\
    2 & kinship & 61-22 & No-Yes & Yes-Yes \\
    3 & kinship & 60-25 & Yes-No & Yes-Yes \\
    4 & Parent-Child & 55-20 & No-No & Yes-Yes \\
    5 & Parent-Child & 60-30 & No-No & Yes-Yes \\
    6 & GP-GC & 78-26 & No-No & No-Yes \\
    7 & kinship & 60-25 & No-No & Yes-Yes \\
    8 & GP-GC & 63-15 & No-No & No-Yes \\
    9 & GP-GC & 68-16 & No-Yes & Yes-Yes \\
    10 & kinship & 61-22 & No-No & Yes-Yes \\
    11 & GP-GC & 62-16 & No-Yes & Yes-Yes \\
    12 & GP-GC & 70-15 & No-No & Yes-Yes \\
    \bottomrule 
\end{tabular}}
\end{table}

\subsection{Procedure}
\begin{figure}[]
\centering
\includegraphics[width=0.95\linewidth]{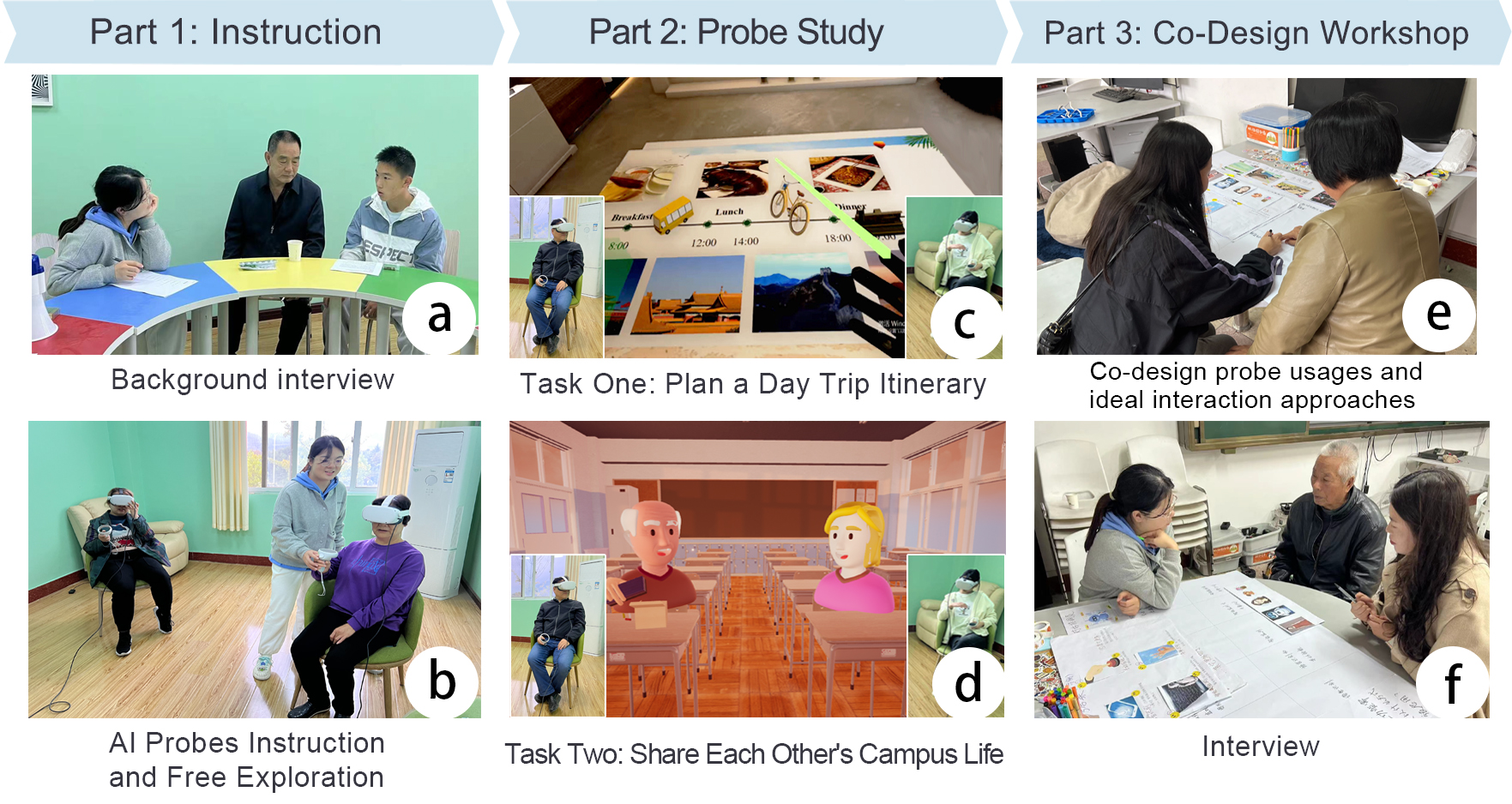}
\vspace{-0.2cm}
\caption{The user study process is structured into distinct parts, each represented by a column. The session began with a background interview (a), followed by a comprehensive tutorial on VR operations and AI probe usage (b). Subsequently, participants engaged in an AI probe study, initially focusing on planning a day's itinerary (c), then shifting to sharing stories about their campus life (d). The session culminated in a co-design session, encompassing both brainstorming activities (e) and thorough in-depth interviews (f).} 
\Description{This figure introduces the process of our user study. We started with an introduction and background interview. They were then familiarized with VR through a tutorial and introduced to three types of AI probes. The study involved two tasks: planning a day itinerary and sharing campus life stories, utilizing the AI probes for inter-generational communication. Finally, we conducted a co-design session with brainstorming and in-depth interviews to explore AI use cases and gather feedback on AI design and functionality.} 
\label{fig:process}
\end{figure}

Our workshop included three parts as shown in \autoref{fig:process}. The session lasted an average of 1.5 hours. The study was conducted in a laboratory setting. Two participants were in the same room, with two moderators present to assist in the experimental process. Before the start of the study, written informed consent was obtained from each participant, with the study purpose and task. All participants were informed of their right to withdraw from the study at any time, although no participants chose to do so. \rv{All participants received a payment of 100 RMB at the end of the study.} \sout{ All participants were given payment at the end of the study. }

\subsubsection{Part 1: Background Interview and Instruction}

Our study began with an introduction of the study and tasks, along with a background information interview (e.g., age, communication frequency, VR experience, and AI experience). We then provided participants with the basic information and tasks about the study. To avoid potential frustration, we also informed them about the limitations of the AI probes, such as the possibility that AI responses might not fully meet their needs and that AI-generated suggestions might not always be correct. 

In the instruction phase, we guided participants to put on the VR headsets, as shown in Figure~\ref{fig:process}-b. Considering that most participants had no prior experience with VR, we started by teaching them the basic operations of VR (e.g., selecting UI elements and manipulating virtual objects).
Next, participants learned to use the AI features one by one. A virtual screen in front of participants displayed the description and instructions for each AI feature in text form in VR (as shown in Figure~\ref{fig:probe}-6). Participants were required to follow these instructions to learn how to use the AI probe and click the ``Learn Next'' button to proceed to the next AI feature.
During the learning session, two moderators observed and provided guidance as needed. Once participants were comfortable using the AI features, we moved them to the formal experimental tasks.


\subsubsection{Part 2: AI Probe Study}
\label{sec:task}
After verifying participants' comprehension of the AI probes and VR manipulation, participants completed the following tasks: 1) Plan a day itinerary and 2) Share each other's campus life. These two tasks represented the two most common activities in inter-generational communication in everyday life: co-negotiation and life-sharing~\cite{lin2002conversation, fuchsberger2021grandparents}. This part was designed to be completed within an approximate timeframe of 30 minutes.

\textbf{Task 1: Plan a day itinerary.} The goal of this task was to stimulate discussion and decision-making between the two participants under a fixed topic. In the task, two participants collaborated to complete a day trip schedule (e.g., choosing and determining meals, attractions, transportation, etc.). They were provided with a blank schedule planner with a timeline marking several points in the day, which they were required to fill in by generating pictures or 3D objects, as shown in \autoref{fig:process}-c. Participants were free to use any of the AI probes to complete the task. 

\textbf{Task 2: Share each other's campus life.} This task encouraged participants to share and exchange stories about experiences unfamiliar to the other generation. Specifically, \sout{two participants} they took turns discussing their campus lives, with interruptions for clarification or engagement allowed to promote interactive dialogue. Participants did not have specific goals to accomplish and could stop the task at any time. They were also free to use AI probes to support their communication during the storytelling process.

\subsubsection{Part 3: Co-design workshop}

Finally, we conducted a co-design session with participants. First, the inter-generational pairs engaged in brainstorming sessions for about fifteen minutes without the presence of the moderators, as shown in Figure~\autoref{fig:process}-e. They were required to discuss how the AI probes supported their communication in VR, the challenges they faced while using the AI probes, and their expectations for future interactions with AI in VR (i.e., preferred interaction strategies and the expected visual representations of AI). We provided paper, colorful sticky notes, markers, and various stickers to help them document their discussions.

After their discussion, two moderators entered the room to conduct in-depth interviews with both participants at the same time to gather detailed insights about the outcomes of their discussion. Our main objective was to encourage the participants to share the reasoning behind their conclusions.

\subsection{Data Analysis}
The qualitative analysis was conducted following the Grounded Theory Approach~\cite{charmaz2006constructing}, aiming to generate a rich and empirical examination of how AI supports inter-generational communication in VR and how participants expect AI to be designed in the future. Following McDonald et al.'s guidelines for qualitative analysis in HCI~\cite{reliabilityNora}, we prioritized identifying recurring concepts and categories over inter-rater reliability.

Specifically, our qualitative analysis contained the following steps:
1) Data Organization. Our data consisted of observational notes and audio recordings of the interviews. The observational notes were taken during the VR experiments, focusing on notable user behaviors and verbal expressions. After completing all experiments, we reviewed the experiment videos to supplement any missing details in the observational notes. The audio recordings were transcribed using DeepL\footnote{https://www.deepl.com} and checked for accuracy by the primary authors;
2) Data Familiarization. All authors thoroughly read the observational notes and participants' narratives to understand common behaviors, perspectives, and expectations regarding AI use in VR;
3) Independent Coding and Category Formation. Each author independently created initial codebooks from three random transcripts~\cite{macqueen1998codebook}. Affinity Diagramming was then used to sort and integrate the coded information into preliminary categories~\cite{beyer1999contextual};
4) Category Refinement. Through iterative meetings, all authors discussed their codes and initial categories to identify discrepancies and make necessary revisions until a consensus was reached;
5) Quote Extraction. One author extracted relevant quotes from all the data based on the refined categories;
6) Further Categorization. The authors further refined the categories and their relationships, using the extracted quotes to create a comprehensive description of how AI supports inter-generational communication in VR.

\section{Findings}

\sout{Our results demonstrate how AI-powered VR could support inter-generational communication among the participants. In this section, we present the key findings of our two RQs.}

\begin{table}[h]
  \caption{Key Findings of RQ1: How AI-powered VR was leveraged to support inter-generational communication. The examples in the table were collected from observations in the study and participants' envisioned applications of AI-powered VR in workshops. Envisioned scenarios were distinguished with an asterisk (*).}
  \label{tab:rq1}
  \renewcommand{\arraystretch}{1.3}
  \scalebox{0.9}{
  \begin{tabular}{p{2.8cm}p{4.2cm}p{6.5cm}}
    \toprule[1.2pt] 
    Key Findings & AI-powered VR Capabilities & Examples  \\
    \midrule 
    \textbf{AI-powered VR enhances mutual understanding} & \textbullet \, Transforming abstract conversation content into concrete forms to better exchange ideas and thoughts & \textbullet \, Participants used Content Generation to generate an image of `chemistry laboratory' to introduce their dream profession; used Info Assistant to explain `involution (nèi juǎn in Chinese)' \\
     & \textbullet \, Providing various ways to interact with generated content to facilitate deeper communication &  \textbullet \, Participants pointed to the roller in the AI-generated picture of the mimeograph machine and explained why the mimeograph machine could dirty the hands; interact with daily tools more intuitively for better imparting life skills to each other* \\
    \textbf{AI-powered VR fosters conversation fluency} & \textbullet \, Providing interventions to avoid awkward silences, maintain balanced conversations, and manage potential friction & \textbullet \, Chat Facilitator suggested \textit{``Could you share something about your closest classmates?''} when the conversation falls into silence \\
     & \textbullet \, Personalizing AI representations to make its interventions more acceptable &  \textbullet \, Participants recommended a realistic, context-appropriate AI appearance to provide convincing guidance for public activities; an inanimate virtual agent to reduce concerns about privacy and authenticity for private conversations; and friendly, cute virtual agents to create a relaxed and joyful atmosphere for casual talk* \\
    \textbf{AI-powered VR promotes active participation} & \textbullet \, Providing content-specific spaces to offer an immersive and engaging conversation experience & \textbullet \, Content Generation switched the VR space to an old classroom when the OGs were reminiscing about their campus experiences \\
     & \textbullet \, Modifying users' avatars or virtual environments based on their emotions to ensure positive communication experiences 
     & \textbullet \, Emotional Visualization could provide a bright and colorful environment in VR when the conversation is sad; offer vibrant and dynamic virtual effects in VR when the conversation becomes dull*  \\
    
    \bottomrule[1.2pt] 
\end{tabular}}

\end{table}

\subsection{RQ1: How could AI-powered VR be leveraged to support inter-generational communication?}

We found that AI-powered VR was leveraged to enhance mutual understanding, foster conversation fluency, and promote active participation. The key findings are summarized in \autoref{tab:rq1}.


\subsubsection{AI-powered VR enhances mutual understanding between two generations} Mutual understanding is important for inter-generational pairs to comprehend each other's thoughts, bridging generational knowledge gaps and enhancing empathy. We outlined two benefits of how AI-powered VR supports mutual understanding. 

First, \textbf{AI-powered VR transforms abstract conversation content into concrete forms, such as images, videos, 3D models, and virtual scenes, enabling different generations to share and understand each other's ideas and thoughts more easily and intuitively.}
During the probe study, participants used Content Generation to create images and 3D objects to introduce concepts that were unfamiliar to another generation. For example, when the older participants shared stories of their past school experiences, all of them tried to generate virtual scenes or objects, such as saying \textit{``Please take us back to my childhood classroom,''} (P4-OG) \sout{or \textit{``Show me an old, dilapidated stool made of wood''} (P2-OG). They then continued telling the story using the generated items or spaces as references.} Similarly, when younger participants described their current jobs or future aspirations, \sout{to their elders,} they often generated images, such as \textit{``a chemistry laboratory''} (P4-YG) or ``a flag bearer raising a flag'' (P1-YG). They indicated that incorporating visual aids would enhance clarity and prevent misunderstandings or confusion.

Additionally, participants found the Info Assistant useful to introduce era-specific vocabulary or events. \sout{, which often reflected particular cultural contexts or trends.} This kind of vocabulary or events may be familiar to one generation but completely unfamiliar to the other. Participants suggested that AI played a crucial role in offering clear and neutral explanations, helping two generations to grasp the meanings and origins of these terms. For instance, P7-YG mentioned the term ``involution (nèi juǎn in Chinese)'' when sharing her college experience, which confused the older listener. To clarify, she employed the Info Assistant to offer a detailed background of the term. \sout{They appreciated the clear and straightforward explanations that AI offered. P7-YG indicated that ``It offered cultural and sociological implications, current relevance, and examples of its use in various contexts, far surpassing my explanations''. }
 \par

Second, \textbf{AI-powered VR provides various ways to interact with generated content, such as pointing, grasping, zooming, and rotating, drawing listeners' attention to specific points and facilitating effective communication between generations.} \sout{In the probe study, participants indicated that AI-powered VR allowed them to interact with AI-generated content, which enabled two generations to engage more deeply in conversation and enhanced communication efficiency.} P7-YG said, 
 \textit{``In traditional phone calls, I find myself not fully engaged because I frequently struggle to understand what he is saying, making it easy for me to lose track of the conversation. \rv{...But in this system, I stay engaged, constantly following the virtual content he shares.} \sout{ However, with the AI-powered VR system, I am actively involved because I need to interact with the virtual content my grandfather is showing me. This shared experience helps us connect more deeply and makes our conversations more meaningful.''} } Additionally, by engaging with generated content, both parties experienced easier presentation and learning. \sout{enhancing inter-generational communication efficiency.} When sharing campus life, P4-OG reminisced about using a mimeograph machine to produce exam papers, noting how it often left her hands dirty. She generated a picture of the \textit{``mimeograph machine''}, grasped the picture, and pointed to the roller, explaining to P4-YG: \textit{``You need to lay the paper flat on the template and then use an ink-filled roller to print it. You constantly have to adjust the template, which can get messy.''} \sout{She expressed that she liked interacting with images because it was difficult to explain using words alone. } \par

In the Co-design Workshop, participants envisioned that they could interact with AI-generated content in VR to impart life skills to each other remotely, enhancing inter-generational knowledge transfer. Compared to 2D screen-based communication, the 3D world of VR provides more realistic object displays and intuitive interactions that closely mimic real-life experiences, making it especially beneficial for older adults. \sout{On the one hand,} Therefore, the older generations could effectively learn how to use the electronic tools from younger generations by generating these tools and practicing in VR. As the P6-OG indicated, \textit{``my daughter often brought me smart home appliances like robotic vacuum cleaners to ease my workload. However, we struggled to use them and often abandoned them when she was not home.''} \sout{They thought that using voice or video calls to ask for help was inefficient and often failed to resolve the issues. However, generated 3D objects in a VR environment could address this gap since they could repeatedly try and practice these skills without the fear of real-world consequences associated with failure.} \rv{They found voice or video calls inefficient for troubleshooting, but VR-generated 3D objects could help bridge this gap, allowing them to practice without real-world consequences.} Similarly, younger participants also indicated their need to seek advice from their elders on life skills, such as \textit{``learning grandma's signature dish''} (P6-YG). They believed that VR Content Generation could effectively aid them in learning life skills hands-on from their elders.

\subsubsection{AI-powered VR fosters conversation fluency between two generations.} \sout{Based on our probe study and workshops,} We identified two ways in which AI-powered VR facilitates conversation fluency. 

\textbf{AI-powered VR could suggest topics or provide timely interventions based on contextual content, helping two generations avoid awkward silences, maintain a balanced conversation, and manage potential friction.} AI could comprehend user dialogues and \textbf{provide potential topics and activities} to avoid awkward silences and keep the conversation flowing. During the study, eight pairs of participants encountered silence during their interactions. For example, after P1-YG stated he was doing well in school, \rv{both he and the OG participant fell silent.} \sout{ he fell silent. P1-OG, too, did not make any further inquiries or continue the conversation.} Sensing this, the Chat Facilitator intervened with a voice suggestion, \textit{``Could you share something about your closest classmates?''} Participants appreciated this mediator \sout{from the Chat Facilitator} and felt it kept the conversation active. \sout{ As P1-YG highlighted, ``When discussing my school life with my grandma, I was unsure about her interests, leading me to briefly touch on basic information before stopping. However, this doesn't mean I lacked the desire to chat with her. The facilitator suggested the potential topics that we could discuss. I appreciated that.''} \rv{Participants envisioned that AI could suggest mutually interesting topics based on their personalities, backgrounds, and previous conversations.} \sout{ participants indicated that their differences in experiences, knowledge, and interests often led to limited topics for discussion and awkward silences in conversation. They hoped that the AI could suggest topics of mutual interest based on their personalities and life backgrounds, combined with the content of their previous conversations. In addition to suggesting potential topics for discussion, P9-YG suggested that AI could also recommend activities for them to engage in next. For instance, watching a movie related to their conversation topics or participating in psychological exercises together could be beneficial.}

Participants perceived that AI could \textbf{balance the participation} of two generations by providing real-time interventions. During the study, AI was activated in seven pairs of participants to balance their conversation. For example, in Task 1, younger participants may independently craft travel plans without seeking advice from their older participants. In this instance, the Chat Facilitator provided suggestions using a voice prompt: \textit{``Let's have both sides discuss together the places you would like to visit.''} They appreciated the reminder of AI and thought it made them attentive to the other's feelings during the conversation. \sout{As P3-YG mentioned, ``I got so wrapped up in making the plans that I forgot to discuss the plans with them. I might have ended up doing the whole schedule by myself if it weren't for the AI reminding me.''} 
Conversely, in Task 2, the older participants often engaged in extended storytelling, ignoring whether the younger listeners were interested. Although the younger participants generally refrained from interrupting the storytelling, believing it to be impolite and disrespectful towards the older generations, they hoped that AI could serve as a mediator to ensure the harmonious progression of the conversation. \sout{``having to listen all the time drains me. It made me feel bored and I found myself zoning out'' (P6-YG). They thought}  The older participants also indicated their welcoming of AI's reminder. As P5-OG described, \textit{``I probably didn't realize this problem because I was excited to tell my story at this time. I'm grateful for the AI's reminder. I think discussing with him together is much more enjoyable than just talking by myself.''} 

Participants also envisioned that AI could \textbf{offer fair interventions to manage potential friction}, maintaining a harmonious atmosphere in communications. During the co-design workshop, \rv{participants suggested using AI as a mediator in daily conversations to mitigate conflicts.} \sout{ participants expressed the idea of employing AI as a mediator in their daily conversations, particularly to help mitigate conflicts.} In this case, AI could intervene in conversations by suggesting a temporary cool-down or offering objective solutions. For instance, in task 2, P8-OG opined that today's campus life is happier than before, but P8-YG countered, citing increased academic pressure now. This led P8-OG to criticize P8-YG for lacking resilience and endurance, causing unpleasantness in their conversation. P8-YG suggested that \textit{``this was a moment where AI could intervene.''} \sout{It could be beneficial to prevent the argument from escalating by offering advice, thereby maintaining a harmonious atmosphere}
Intriguingly, several participants expected they might prefer AI over human coordinators for resolving inter-generational conflicts. This preference stemmed from their perception of AI as being neutral and unbiased, unlike humans who could be influenced by emotional attachments or social connections. For instance, P8-YG remarked, \rv{\textit{``When I have conflicts with my grandfather, I prefer listening to AI over my parents or grandmother, as they tend to take sides and can't mediate impartially.''} This highlighted the perceived impartiality of AI in conflict resolution.} \sout{ ``When I have conflicts with my grandfather, I'd rather listen to AI than my parents or grandmother because my parents always side with my grandfather, and my grandmother with me. I feel that none of them can persuade us rationally and impartially.'' This comment underscored the perceived impartiality of AI in conflict resolution. }

\begin{figure}[]
\centering
\includegraphics[width=0.7\linewidth]{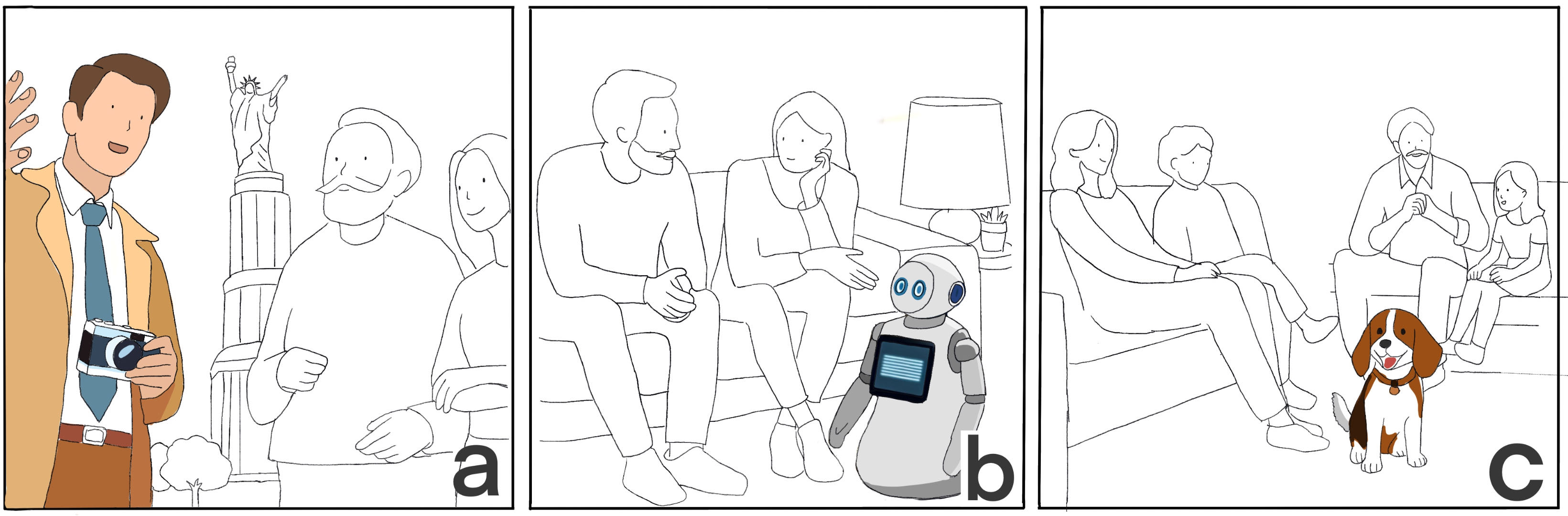}
\vspace{-0.2cm}
\caption{Illustration of participants' suggestions for AI's visual representation: a) shows a realistic, context-appropriate AI appearance such as a tour guide describing the attraction. Participants indicated that such representations would be especially beneficial in VR-shared activities, enhancing immersion and providing convincing guidance; b) depicts an inanimate virtual agent as a robot. Participants believed that it reduces their concerns about privacy and authenticity, making them more suitable for private conversations; c) represents a friendly and cute pet as a virtual agent. Participants believed that such representations were effective in easing tension and adding humor to interactions.} 
\Description{} 
\label{fig:visualAI}
\end{figure}

\textbf{AI-powered VR provides AI with embodied and personalized avatars that adapt to different communication contexts, making AI's intervention more acceptable for users.} 
\rv{In the workshop, participants noted that while they found verbal interventions helpful for facilitating communication, a visual representation of AI would increase their trust, as it seemed more human-like and reliable.}
\sout{In the workshop, participants pointed out that while they believed verbal interventions provided by the system helped facilitate communication, they felt that these interventions alone were insufficient to build trust and rapport. 
Participants believed they would be more likely to trust AI when it has a visual representation because it appeared more human-like and reliable.}
P5-OG emphasized, \textit{``In Task 1, without an embodied avatar, the AI's suggestions felt as if they suddenly emerged from the air, which seemed abrupt and made us hesitate in accepting the suggestions.}
\rv{The embodied avatar allowed users to feel they were interacting with a real person rather than an intangible system. \sout{enhancing their immersion and engagement.} Moreover, avatar appearances can be customized according to user needs and interaction scenarios, enabling AI to integrate seamlessly into conversations and adapt dynamically.}
\sout{ The embodied avatar made users feel as though they were interacting with a real person rather than just an intangible system. This enhanced their sense of immersion and engagement. Additionally, these avatar appearances can be customized and contextualized according to user needs and interaction scenarios, enabling AI to integrate more seamlessly into conversations and dynamically adapt to various communication contexts. }
Participants proposed three distinct types of visual representations, each suited to different scenarios, as shown in Figure~\autoref{fig:visualAI}. \par

First, participants suggested \textbf{a realistic, context-appropriate AI appearance}, such as a virtual tour guide or a teacher. \rv{They believed such representations would enhance immersion and provide convincing guidance in VR-shared activities.} \sout{Participants believed that such representations would be particularly effective in VR-shared activities, enhancing user immersion and providing convincing guidance.}As P4-YG noted: \textit{``A guide dressed in formal attire in a VR museum is more convincing than a dog or cat.''} However, this realistic avatar\sout{despite its effectiveness in providing immersive experiences for users,} may hinder casual inter-generational conversation due to its strong and invasive presence, as P6-YG commented: \textit{``Such a strong presence of AI feels like it's listening in on everything we say, which is somewhat uncomfortable.''} 

The second proposed form was \textbf{an inanimate virtual agent}, such as a robot. Participants indicated that, in contrast to lifelike avatars, the non-human appearance of robots reduced their concerns about privacy and authenticity, making them more suitable for private conversations. Additionally, robots were seen as impartial, free from emotions or biases, and thus, ideal for conflict mediation. As P3-YG highlighted \textit{``I feel more open to listening to the unbiased advice from the emotionless robots.''}

Lastly, participants suggested \textbf{friendly and cute virtual agents}, like cartoon characters or pets, for their ability to create a relaxed and joyful atmosphere. This approach may mediate conversational dynamics, especially during awkward or silent moments between generations. These avatars were seen as effective in easing tension and adding humor to interactions. As P1-YG remarked, \textit{``I prefer cartoon puppies as they feel like part of the family and their interactions are natural and unforced.''} \sout{Their cuteness makes me more receptive to their advice, making cartoon characters a more comfortable choice for receiving guidance.}

\subsubsection{AI-powered VR promotes active participation between two generations.} We identified two reasons why AI-powered VR made inter-generational communication more active.

\textbf{AI-powered VR provided content-specific immersive environments and engaged both participants in active conversation.} Participants commended the benefits of switching the virtual scenes in VR according to their communication content. They thought that it provided an immersive communication experience that fostered a greater willingness to listen and share stories. 
Participants indicated that their current storytelling process was usually led by the older generations while the younger generations often lacked a sense of involvement and empathy. As P10-YG mentioned \textit{``Sometimes, I get impatient listening to her youth stories because they're hard for me to relate to... However, I found the stories more engaging during VR experiences, as they make me feel like I'm reliving those moments with her.''} They appreciated that AI could generate scenes in VR (e.g., an old family home or a meaningful location from the past) or 3D virtual items (e.g., agricultural tools or stroller) to enhance the storytelling experience and actively engage listeners. \sout{in the storytelling process.} 

Additionally, participants envisioned that AI could provide active co-learning experiences by creating historical scenes and geographical landscapes mentioned in textbooks within VR environments. They indicated that co-learning was a common activity they engaged in at home, and AI could potentially benefit this activity by boosting learning interests and outcomes. For example, the P12-OG expressed that ``AI can help bring to life the scenes described in ancient poetry, allowing us to immerse ourselves in these landscapes witnessed by poets centuries ago in VR.'' 
\sout{Participants also suggested that AI could help them automatically ``dress up to resemble a local person from that era.'' (P12-YG) to make them feel more engaged during the co-learning process. }

\textbf{AI-powered VR can modify the users' avatars or virtual environments based on their emotions to ensure positive communication experiences.} During the VR experience, participants' pronounced emotions were visualized in real-time. \sout{(e.g., rainy effects for negative emotions and pink bubbles for positive emotions). }

Participants appreciated the visualization of their positive emotions. P7-YG participant expressed, \textit{``It made me feel my uncle's joy, which in turn made me happier and more willing to discuss the topic further with him''}.
Participants suggested two ways AI could help express their happiness. The first way is to automatically modify the expression of users' avatars, such as \textit{``showing hearty eyes, wide smiles, or even playful winks''}. Embodied avatars, as virtual representations of users, can effectively influence the social experience during interactions~\cite{wei2022communication}. Participants believed that this subtle intervention could create a pleasant communication atmosphere without disrupting the flow of the conversation.
The second approach is to modify VR environments to exaggerate users' positive emotions, such as \textit{``providing joyful background music, altering the surrounding environment, and making objects around them dance''}. Participants felt that this visual enhancement of positive emotions could effectively increase their emotional involvement with each other. They believed that immersive VR could make these enhancements more impactful, providing a more realistic experience than screen-based communication platforms.

While most participants saw value in visualizing positive emotions to enhance the conversation atmosphere, they unanimously preferred not to exaggerate or display their negative emotions to others. As P9-OG mentioned, \sout{We don't want the VR world to show our bad moods.} \textit{``Imagine feeling down and suddenly there's a storm in the VR --- that'd just make us feel worse.''} They suggested that if the conversation is serious or sad, the scene could transform into something bright and colorful, like a sunny park or a lively beach, to lift spirits and re-energize the dialogue.  \sout{I think it's better if the VR environment could help us chill out and get our mood back on track when we're feeling low. If the conversation becomes inactive or dull, the setting might shift to something more vibrant and dynamic, Therefore, they could better engage in positive communication and maintain pleasure experiences.}

\subsection{RQ2: What are the challenges of using AI to support VR-based inter-generational communication?}
\label{sec:rq2}

\sout{Based on the probe study and co-design workshop,} We identified three challenges participants faced when using AI to support their communication in VR, including the inappropriate timing and manner of AI interventions in VR, the difficulties of AI in comprehending participants' intentions, and participants' over-reliance on AI. 


\subsubsection{Participants felt that the inappropriate timing and manner of AI interventions in VR disrupted the conversation flow and disturbed the harmonious conversational atmosphere.}
Participants perceived that VR's immersive environment makes public interventions more prominent and effective in highlighting issues. Consequently, AI's public interventions in highlighting others' problems become more inappropriate and impolite, particularly when involving the elderly. As P9-YG expressed, \textit{``We need to be respectful when talking with elders. Even though AI's public intervention stopped my grandfather from going on and on, it made the conversation a bit awkward afterward.''} Additionally, participants reported that AI sometimes intervened at inappropriate times during their conversations, disrupting the natural flow and causing discomfort. P12-YG indicated, \textit{``While my grandparent sharing his school story, the AI suddenly interrupted. We were all engaged and didn't need any assistance.''} \sout{, so this interruption felt unnecessary and rude.}

\begin{figure}[]
\centering
\includegraphics[width=1\linewidth]{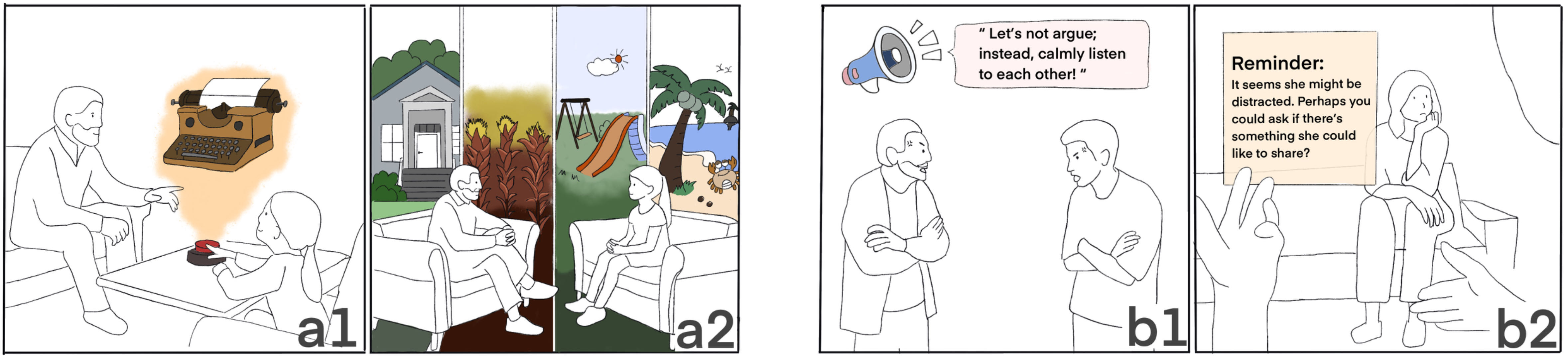}
\vspace{-0.2cm}
\caption{Illustrations of participant-proposed AI interaction strategies in VR. Two activation methods were envisioned on the left: a1) passive activation, which emphasizes user autonomy and aims to minimize interruptions; and a2) proactive activation, where the virtual environment in VR dynamically changes in response to the user's emotions or conversation content. Two intervention strategies are proposed on the right: b1) prompts visible to both parties involved in the communication; b2) private prompts designed to discreetly remind the user of their behavior.} 
\Description{} 
\label{fig:wakeup}
\end{figure}

Participants suggested that future designers could make AI interventions more appropriate and acceptable in VR. They envisioned two AI activation approaches and two AI intervention forms, as shown in \autoref{fig:wakeup}. They also outlined how these could be applied in inter-generational interactions.

Generally, participants favored a manual, user-controlled \textbf{passive method} to invoke AI's features for most scenarios (\autoref{fig:wakeup}-a1), which enables users to engage the AI as needed through buttons, voice commands, or gestures. Such control is essential for maintaining user autonomy and minimizing interruptions, especially during private or in-depth conversations. \sout{As P8-YG mentioned: ``It would be too dazzling if AI generated a scene wherever I was chatting. It made me feel like we were creating rather than chatting.''}
Nevertheless, participants saw the potential for a \textbf{proactive method} for specific situations like emotional downturns (\autoref{fig:wakeup}-a2). They noted that during intense emotions such as sadness or anger, they tended to become absorbed in their feelings, focusing on the emotions rather than seeking ways to alleviate or manage them. In these situations, participants appreciated the idea of proactive AI intervention. P11-OG stated that: \textit{``When I become angry or sad during a conversation, I hope VR can display beautiful scenery to help regulate our emotions and improve the relationship.''}

In terms of intervention manners, participants believed \textbf{publicly displayed} intervention could be beneficial for providing objective solutions and information to facilitate communication, as they can help clarify misunderstandings and ensure that both parties are on the same page. For example, if a misunderstanding leads to an argument, AI could intervene with prompts noticeable to both parties, such as an audio announcement to suggest a cool-down period or provide an objective solution. \sout{As P11-YG mentioned, ``When conflicts arise, we may both be irrational and may even say something hurtful to the other party''. They believed that this direct and proactive strategy could effectively modify the communication atmosphere, prevent emotional escalation, and avert conflict deterioration.}
Additionally, participants suggested that the \textbf{private prompts} is more acceptable and polite to remind them in situations where one party dominates the conversation or says something that makes the other person feel uncomfortable (\autoref{fig:wakeup}-b2), especially for the elderly. For instance, a discreet text prompt such as \textit{``You've spoken at length; let the other person share their views''} could be displayed only in the dominant speaker's VR environment; Similarly, when one party discusses uninteresting topics for the other, a prompt like \textit{``It seems that they don't want to talk about this. Try asking about their enjoyable moments at school.''} could be shown to the speaker.\sout{to encourage positive communication experiences.} Participants indicated that the private prompts, compared to cues visible to both parties, would maintain a positive conversation atmosphere and flow while preserving the dignity of the prompted user. \sout{Younger participants found this particularly valuable, as they might be reluctant to directly interrupt or express discomfort to older conversation partners, such as when the elderly overly dominated the conversation or broached unwelcome topics. P10-YG shared that ``there are times when elders bring up uncomfortable topics, like questions about my boyfriend or salary. I'd prefer the AI to subtly indicate my discomfort, rather than having to voice it myself. Visible prompts to both participants could be awkward in this scenario.''} Older participants also saw value in private prompts, acknowledging that they sometimes did not realize the younger generation's disinterest in their topics. P9-OG stated, \textit{``If I'm speaking excitedly, I might not realize that others aren't getting a chance to speak. I hope the AI can remind me discreetly.''}

\subsubsection{Participants found that AI sometimes misunderstands their intentions and provides wrong responses, impacting inter-generational communication efficiency and potentially heightening conflicts.} 
\sout{Through the observation of the probe study and the analysis of the workshop outcomes,} we attributed this challenge to two factors:

Firstly, \textbf{the lack of non-verbal cues} led to failures in understanding participants' intentions in VR. Compared to screen-based communication tools, users are more likely to express non-verbal cues --- such as gestures, facial expressions, head orientation, gaze direction, and proximity --- when communicating with others in VR~\cite{smith2018communication, abdullah2021videoconference}. In the study, we observed that participants unconsciously employed non-verbal cues during their interactions with the AI. For example, P4-OG pointed at a picture in VR and asked the AI, \textit{``Which section of the Great Wall is this?"} The AI was unable to answer the question because it did not recognize the participants' pointing gestures and pointing areas. Participants expressed a desire to integrate multi-modal user inputs in VR, such as body language, facial expressions, gestures, gaze, and interactions, to better perceive their intention when communicating with AI.
 \sout{However, in our system, we did not track and provide users' non-verbal cues to the AI, resulting in AI missing valuable information that could convey participants' intentions. This limitation led to the AI being unable to provide correct responses.}

Secondly, AI's \textbf{limited understanding of participants' background}, including familial dynamics, communication habits, and regional language, led to conversation disruptions and may even exacerbate conflicts. Participants acknowledged the complexities inherent in each family, posing a substantial challenge for AI in accurately understanding conflicts and responding suitably. Consequently, they worried that AI might intervene inappropriately in situations where it should not. For example, P9-YG mentioned, \textit{``I think it is important for AI to understand that my grandfather is accustomed to speaking loudly due to hearing loss, which should not be interpreted as a sign of anger.''} Additionally, participants voiced concern that AI might make erroneous interventions because it may not fully grasp the dynamics of conflicts. As P11-YG mentioned, \textit{``Sometimes, a simple remark from my dad can make me angry. This is because there are long-standing sibling rivalries or past incidents that substantially impact the dispute. I worry that AI, lacking family background, might not understand the root of my anger and provide effective advice or solutions.''}  
Participants also highlighted the AI's inability to recognize dialects impairing their communication experience. This issue was particularly evident in China, where many older adults, especially those residing in rural areas, struggled with fluency in standard Mandarin. Our study revealed a strong preference among most participants for using local dialects when communicating with family members. However, to facilitate better understanding by the AI during the study, they were compelled to switch from their native dialects to standard Mandarin, which disrupted the natural flow of conversation. \sout{This switch was perceived as a distraction, detracting from the natural flow of conversation. As P7-OG mentioned, ``At home, I typically communicate with my granddaughter in our dialect. However, I have to use Mandarin when giving a prompt, which made me feel awkward and affected my active participation..''}


\subsubsection{Participants' over-reliance and perceived fairness in AI led to negative consequences.} 

During the experiment, we found that participants sometimes over-trusted the results provided by AI, leading to unreasonable decisions. For instance, when P6 requested a one-day tour plan for Beijing, the AI provided an extremely packed and overwhelming schedule. The suggested itinerary included visiting the Forbidden City, Tiananmen Square, and the Great Wall in the morning, followed by the Summer Palace, Beihai Park, and the Beijing Film Museum in the afternoon. Despite the impracticality, three groups of participants accepted the AI's suggestions and incorporated these activities into their plan.

Furthermore, during the interviews, two groups of participants mentioned that they perceived AI as fair and unbiased, unlike humans who could be influenced by emotional attachments or social connections. Consequently, they were more inclined to trust AI interventions over those from human facilitators. However, previous research indicates that AI-generated results can contain social biases~\cite{roselli2019managing, liu2022trustworthy}. Without critical judgment of the AI's outputs, this could lead to negative consequences, such as reinforcing existing biases or making flawed decisions.

\section{Discussion}

In this section, we further discuss our findings along the following aspects: 1) the advantages of leveraging AI in VR to enhance inter-generational communication and how this integration addressed issues of inter-generational communication, 2) the challenges of integrating AI to support inter-generational communication in VR and associated design implications, and 3) a comparison of AI facilitators with human facilitators to understand their respective strengths, weaknesses, and suitability in different contexts.

\subsection{The integration of AI and VR to support inter-generational communication}

This work aimed to understand how inter-generational pairs perceive and leverage AI to support communication in VR (RQ1). Previous works have demonstrated the advantages of VR in facilitating inter-generational communication, such as achieving a degree of co-presence and companionship by immersive shared environment~\cite{wei2023bridging, afifi2022using}, cultivating empathy by embodying avatars with each other's appearance~\cite{oh2016virtually, shenxxxxxxxx}, or providing a casual and relaxed atmosphere by engaging in various activities~\cite{wei2023bridging, hoeg2021buddy}. This work extends prior work by leveraging AI to create a \textit{content-aware VR environment} to support a dynamic communication process. Through a probe-based participatory design workshop, we identified three key ways in which inter-generational users wish to leverage AI in VR to support communication. 

First, participants recognized the potential of the integration of AI and VR to \textit{enhance mutual understanding} by bridging knowledge gaps between generations. Prior research highlights a widespread desire to maintain familial connections and share life experiences with family members~\cite{mynatt2001digital, tee2009exploring, neustaedter2006interpersonal}. However, the diverse cultural backgrounds, characterized by unique values, beliefs, and communication styles, often lead to misunderstandings between generations~\cite{mynatt2001digital, tee2009exploring}. Our findings suggest that AI and VR could facilitate \textit{mutual understanding} by allowing users to switch abstract conversation content into concrete objects to effectively exchange their ideas. Additionally, users could further interact with the AI-generated outcomes to discuss and clarify their thoughts. Such interaction could not only address the knowledge gaps but also enrich the life-sharing and storytelling experience, making it more vivid and tangible. 

Second, the integration of AI and VR was seen as pivotal in \textit{enhancing conversation fluency} of two generations through appropriate and real-time interventions. Typically, non-fluent conversations between two generations could result in brief and unsatisfying interactions, limiting opportunities for meaningful knowledge and cultural exchange~\cite{lin2002conversation, wei2023bridging}. Our research demonstrates AI's effectiveness in avoiding awkward silences, maintaining balanced conversations, and alleviating potential conflicts. When integrated with VR environments, AI can embody various visual representations (e.g., teacher, robot, puppy) based on the conversation context, making it easier for users to accept AI's suggestions. 

Third, participants suggested that the integration of AI and VR could \textit{promote active participation} between generations. Prior research suggests that the lack of shared knowledge and the subtlety of emotional expression hinder active engagement across generations, leading to miscommunication, a lack of genuine connection, and a sense of impatience among the youth~\cite{kincaid2013communication, williams1996intergenerational, giles2003intergenerational, peterson2019spectrality}.  
Participants noted that AI could modify the VR environment in real-time based on the context content and emotional tone of conversations, encouraging users to share and listen more attentively to each other, thus enriching the quality of the interaction overall.

Overall, our findings indicate that integrating AI with VR opens new avenues for enhancing communication between generations, providing a promising direction for future studies that aim to use VR in supporting inter-generational communication.




\subsection{Design implications for AI to better support inter-generational communication in VR}
This work revealed the challenges of using AI in VR to support inter-generational communication (RQ2), including inappropriate timing and manner of AI interventions in VR, AI's inaccurate understanding of user intentions, and the issues arising from over-reliance and trust in AI. Based on these user-identified challenges, we outlined key design implications \sout{and considerations} to guide future researchers and developers in creating AI-based VR platforms that enhance inter-generational communication.
Among these design implications (DIs), DI1-4 focuses on how to effectively combine the advantages of AI and VR to enhance inter-generational communication experiences. Meanwhile, DI5-7 are key recommendations derived from our study for using AI to support inter-generational communication, which may apply to VR and other communication platforms.

\subsubsection{Exploring various VR designs for providing appropriate AI interventions to support inter-generational communication}


\sout{After experiencing fundamental functionalities,} While participants appreciated the benefits of integrating AI and VR, they also highlighted issues with AI's inappropriate intervention.\sout{and suggested that future designers should better leverage VR's capabilities to create more appropriate and comfortable AI interactions} Based on participant feedback and suggestions, we identify three promising directions for future work. \sout{integrating AI into VR to enhance inter-generational communication. }

First, we suggest that \textbf{DI1: AI could adapt different representations in VR based on conversational contexts, thereby creating a suitable communication atmosphere and making their suggestions more acceptable.} Similar to users' avatars that can influence others' perceptions~\cite{kolesnichenko2019understanding, mcveigh2019shaping, latoschik2017effect}, in our study, participants indicated that AI's visual representations in VR could also affect their perceptions of AI's interventions. They recommended three kinds of AI representations for different communication contexts: a realistic, context-appropriate AI appearance to provide convincing guidance for public activities; an inanimate virtual agent to reduce concerns about privacy and authenticity for private conversations; and friendly, cute virtual agents to create a relaxed and joyful atmosphere for casual talk. This aligns with HCI research on multi-agents, emphasizing personalized interaction and adaptability in social contexts~\cite{jiang2023communitybots, cardoso2021review}. 
 
Second, we suggest that \textbf{DI2: AI's intervention could be enriched using various VR visual effects, creating engaging and appealing social experiences}. Prior studies have demonstrated the effectiveness of diverse visual representations — such as emotes, particles, creatures, fur, skeuomorphic objects, ambient light, and halos — in conveying users' social behaviors and emotional states in VR~\cite{bernal2017emotional, lee2022understanding, dey2017effects, salminen2018bio}. They show that diverse VR visual effects could make interactions more active and fun, potentially leading to positive emotional states and a greater willingness to communicate~\cite{moustafa2018longitudinal, wei2022communication, roth2018beyond}. These novel approaches offer exciting possibilities for future research. \sout{in enhancing inter-generational communication through the integration of AI in VR environments.}

 Finally, we suggest that \textbf{DI3: future research should leverage appropriate forms of AI intervention in VR (i.e., public or private prompts, and automatic or manual activation) to maintain the smoothness of communication}. In our study, participants found that the public interventions of AI sometimes disrupted the flow of communication and could potentially lead to a negative atmosphere. \sout{Participants suggested the possibility of AI providing private prompts instead. Therefore,} They thought that prompts visible to both parties could be beneficial for providing objective solutions and information. \sout{to facilitate communication as they can help clarify misunderstandings and ensure that both parties are on the same page.} On the other hand, private prompts are suitable for discreetly reminding one user during awkward silences or imbalances in the conversation, thus preventing potential embarrassment or discomfort. Selecting the appropriate form of AI intervention in VR helps maintain a positive conversation atmosphere and flow, ensures that interactions remain engaging, and preserves the dignity and comfort of the prompted user.

 \sout{Overall, we recommend that future research explore more practical and innovative ways to integrate AI into VR environments, building on the potential shown in this study for improving inter-generational communication. }

\subsubsection{Enhancing AI's interpretation of user needs in VR} 

As mentioned in Section~\ref{sec:rq2}, participants often interacted with the AI using non-verbal cues in VR. However, AI failed to understand the users' non-verbal cues correctly, thereby providing incorrect responses. This insight led us to propose \textbf{DI4: AI needs to incorporate users' non-verbal cues, such as postures, gestures, orientations, finger-pointing, and gaze direction, to understand user intentions precisely.} VR can easily track multiple channels of user non-verbal information in VR since users wear headsets and hold controllers~\cite{vinnikov2017gaze, tabbaa2021vreed, luro2019comparative}. By incorporating these non-verbal cues, AI can better understand user intentions and provide more accurate responses. 

Additionally, we propose \textbf{DI5: AI needs to understand the communication context and dynamics among users to better interpret their needs and provide appropriate responses.} 
Communication within families is often more casual and centered around personal topics, a relaxed tone, and language reflecting familial backgrounds~\cite{lin2002conversation, heshmat2021family}. However, participants reported difficulties with AI comprehending this informal style.\sout{especially when dialects, regional languages, or family-specific references were used.} They voiced concerns that the AI may misunderstand family-specific expressions or humor, resulting in confusion and disrupting the flow of conversation. Therefore, it is important for AI to grasp the context and subtleties of users' conversations. To achieve this, future AI designers should ask users to provide basic information during the initial setup phase, such as educational backgrounds, cultures, and common languages. Additionally, for ongoing effectiveness and respectfulness, AI should continuously learn and update its knowledge and algorithms to stay in sync with new information, evolving social norms, and family-specific events~\cite{bansal2019beyond, xu2023transitioning, zheng2022ux}. This detailed understanding would enable AI to grasp the context and nuances of users' speech more accurately. This not only helps AI provide personalized responses that resonate with each user's unique experiences and viewpoints but also helps AI determine appropriate timing for intervention, thereby avoiding interruptions to the natural flow of inter-generational communication.



\sout{By leveraging these diverse inputs, AI can more effectively understand users' requirements and generate outputs that align more closely with their expectations and needs.
}

\subsubsection{Ensuring trustworthy and unbiased AI in inter-generational communication} 

In this study, we observed that some participants placed substantial trust in the information provided by AI, viewing it as a fair and unbiased \sout{inter-generational communication} facilitator. However, based on previous research, we recognize that such trust may lead to certain issues for two primary reasons. Firstly, the AI's outputs are not always accurate~\cite{kim2023chatgpt}. Blind acceptance of AI-generated results can mislead conversations and lead to erroneous decisions~\cite{kim2023chatgpt, fu2024text}. Secondly, AI is not unbiased~\cite{roselli2019managing, liu2022trustworthy}. LLMs used in AI are trained on extensive datasets that often contain inherent social biases~\cite{dehkharghanian2023biased, ntoutsi2020bias, daneshjou2021lack}. These biases can inadvertently influence the AI's behavior and outputs, leading to unintended consequences. For instance, an AI designed to facilitate conversation might unintentionally favor certain cultural norms or gender stereotypes embedded in its training data, thus affecting the quality and fairness of its interactions~\cite{nadeem2020gender}. Therefore, if users unconditionally trust AI's results, not only can it lead to inefficient communication but also reinforce stereotypes or create friction in conversations, particularly when different generations have varying levels of digital literacy and cultural perspectives.
 
We propose two design considerations for future AI designers to mitigate these issues. First, \textbf{users should be reminded to approach AI-generated results with a critical mindset during their interactions (DI6)}. Several strategies could be employed to achieve this. Initially, designers could set clear expectations that AI-provided answers may not always be accurate before users engage with the system~\cite{roselli2019managing}. Additionally, interactive tutorials can be used to educate users on the AI's strengths and limitations, including examples of accurate and inaccurate responses. Throughout the interaction, continuous reminders of the AI's limitations can be provided through periodic notifications or prompts~\cite{aggarwal2023artificial}. For example, after delivering a response, the AI could include a brief message suggesting that users verify the information independently or consider alternative viewpoints. Another approach is to incorporate a feature that highlights the AI's confidence level in its responses~\cite{bruzzese2020effect}. This could be represented as a confidence score or a visual gauge that indicates how certain the AI is about the information provided. 
 \sout{By implementing these strategies, designers can ensure that users are well-informed about the potential inaccuracies of AI-generated answers and are better equipped to use AI as a supportive tool rather than a definitive source of information.}

Second, \textbf{it is essential to implement bias detection and correction mechanisms within the AI system (DI7)}. Previous studies have demonstrated potential ways of identifying and mitigating biases within AI systems. For instance, regular audits of the AI's outputs, inclusive training datasets, and user feedback loops are vital for ensuring a more balanced and fair interaction environment~\cite{liang2022advances, mittermaier2023bias}. Additionally, periodic reviews can identify and rectify biases in the AI's responses, ensuring they remain accurate and impartial. Using diverse and representative training datasets can help mitigate embedded social biases~\cite{sen2020towards, liang2022advances, yang2020predicting}, while real-time user feedback can provide immediate insights into potential issues, allowing for swift adjustments. Mechanisms such as feedback buttons or surveys can empower users to report biased or inappropriate responses, ensuring ongoing refinement of the AI~\cite{schulenberg2023towards}. Addressing these biases is important for the development of practical and meaningful AI-based VR platforms that genuinely enhance inter-generational communication without perpetuating existing societal biases. \sout{By implementing these strategies, AI systems can become more equitable and effective tools for fostering understanding and collaboration across different generations and cultural backgrounds.}

\subsection{Comparing AI+VR with co-located human facilitators in remote inter-generational communication}

Previous research indicates that remote inter-generational communication sometimes relies on co-located human facilitators to coordinate conversation~\cite{jin2023socio, gan2020connecting, yarosh2013almost, raffle2011hello, heshmat2020familystories}. These human facilitators, who might be parents, siblings, or grandparents, frequently manage three-party communication via phone or video calls to facilitate inter-generational exchanges. Such facilitators are often necessary in situations where, for example, a grandchild may not be very familiar with their grandparent living apart, leading to potential awkwardness and a lack of topics to discuss if they converse directly. In these scenarios, parents often act as facilitators to join conversations with their children and grandparents. They offer assistance including providing context~\cite{gan2020connecting}, bridging generational understanding gaps~\cite{jin2023socio}, engaging both generations' attention~\cite{gan2020connecting, raffle2011hello}, and suggesting topics to initiate conversation~\cite{jin2023socio, yarosh2013almost}. These studies show that human facilitators help maintain effective inter-generational communication, enhance mutual understanding between generations, and strengthen family bonds. However, they also highlight several drawbacks associated with human facilitators. For instance, facilitators may sometimes dominate conversations, steering them toward unwanted topics~\cite{jin2023socio, yarosh2013almost}. Additionally, inter-generational pairs may find it uncomfortable to discuss specific topics or personal matters in the presence of the facilitators~\cite{jin2023socio, gan2020connecting}.

Our findings suggest that AI-empowered VR can effectively complement or enhance the coordination offered by human facilitators, such as facilitating mutual understanding, enhancing engagement, and initiating discussions. Furthermore, AI facilitators in VR can mitigate some drawbacks associated with human facilitators~\cite{jin2023socio, yarosh2013almost, gan2020connecting}. For example, by adjusting the level of AI intervention, AI facilitators could avoid dominating conversations, ensuring a comfortable interaction experience. Additionally, the AI can adopt different virtual representations to minimize its presence to help users feel more at ease when discussing private matters. Research also indicates that VR's immersive environment and embodied avatars offer a greater sense of companionship and togetherness, alleviating feelings of longing between generations~\cite{wei2023bridging}.

However, we also found that AI has shortcomings compared to human facilitators. Due to a limited understanding of users' personalities and family histories, AI might fail to detect issues in conversations accurately, potentially disrupting conversations~\cite{maier2022comparing, yarosh2013almost}. Additionally, AI may lack flexibility when handling unexpected situations, especially those beyond its training scope~\cite{yang2020predicting}. Human facilitators, on the other hand, exhibit high flexibility and adaptability, improvising in response to sudden changes and complex scenarios. Moreover, AI facilitators may raise privacy concerns~\cite{zheng2022ux, manheim2019artificial}. Users might be wary of how their data is used, affecting trust. Human facilitators can build trust through direct interaction and personal connection, providing clearer privacy assurances and reducing concerns about data security. Furthermore, AI facilitators cannot update each other's status~\cite{gan2020connecting, jin2023socio} or coordinate varied family schedules as effectively as human facilitators~\cite{heshmat2020familystories, yarosh2013almost}. By addressing these areas, future research can contribute to optimizing the roles of AI facilitators in enhancing inter-generational communication, ensuring that they complement and enhance the benefits provided by human facilitators while minimizing potential drawbacks.

\section{Limitations and Future Work}

We acknowledge five limitations that motivate potential future research directions:

First, our study was conducted with inter-generational pairs from Chinese communities, which might introduce cultural and regional biases to our results~\cite{williams1997young, zhang2001harmonies}. For instance, the optimal ways for AI intervention in inter-generational conversations might vary across different countries. While our findings highlight general AI perceptions and implications in this context, further research with a diverse international sample is crucial to understanding the varying impacts of cultural differences on these perceptions.

Second, this work mainly focused on understanding users' perceptions and needs of AI in enhancing inter-generational communication. We did not conduct comparative evaluations to gauge AI's effectiveness against a standard task completion baseline. Building on our findings, future research should explore the impact of AI-powered VR applications more thoroughly. This could involve using control groups in subsequent studies for a clearer understanding of how AI interventions specifically aid inter-generational communication. 

Third, to prevent long rendering times from disrupting user communication, we employed pre-generated content instead of real-time 3D models and virtual scenes in our study. While this method allowed for controlled and consistent experimental conditions, it limited our ability to explore certain characteristics of AI such as unpredictability and latency in the context of inter-generational communication. Future research should explore the challenges and impacts of using real-time AI generation in VR to gain a more comprehensive understanding.

\rv{Fourth, to ensure consistent experimental conditions and minimize distractions related to avatar appearance, we did not employ an avatar customization function in our study. While this decision facilitated a more focused interaction with AI-powered VR, it restricted our investigation into how discrepancies between user and avatar appearance could affect the sense of embodiment and overall immersive experience. Future research should consider incorporating customizable avatars to examine their influence on user engagement and the dynamics of inter-generational communication.}


Moreover, although our AI probe-based co-design workshop revealed interesting insights, it lasted for a relatively short period and was conducted in a controlled laboratory setting. These factors may compromise the authenticity and practical applicability of the experimental outcomes~\cite{morgado2017scale}. Future research should focus on longer, more extensive, and ecologically valid studies outside the lab to better understand AI's long-term effects and gather practical insights for daily use in inter-generational communication.

\section{Conclusion}
Inter-generational communication \sout{involves the exchange of experiences and perspectives between different generations within a family,} is crucial for enhancing family bonds. 
\rv{Through a probe-based participatory design study with twelve inter-generational pairs, our work examines how AI can enhance inter-generational communication in VR by improving mutual understanding, conversation flow, and participation. It also highlights challenges, such as inappropriate AI interventions, misinterpretation of user intentions, and over-reliance on AI. We provide design implications to address these issues, aiming to develop AI tools that are both technologically advanced and sensitive to emotional and family dynamics, fostering stronger family connections.}
\sout{Our study explores the significant role of AI in enhancing inter-generational communication within VR environments, including enhancing mutual understanding, improving conversational fluency, and encouraging active participation, highlighting the positive impacts of AI and VR integration in bridging generational gaps in VR interactions. Meanwhile, our participatory design approach reveals the challenges 
of using AI to support VR-based inter-generational communication, including inappropriate timing and manner of AI interventions in VR, AI's inaccurate understanding of user intentions, and the issues arising from over-reliance and trust in AI. Based on these challenges, we outlined some key design implications to guide future researchers and developers. This research paves the way for future developments in communication tools, emphasizing the importance of AI which is not only technologically advanced but also sensitive to the complexities of human emotions and family dynamics, offering new opportunities for strengthening familial ties and improving the well-being of all generations.}

\bibliographystyle{ACM-Reference-Format}
\bibliography{sample-base}

\appendix

\section{Appendix}

\begin{table}[h]
\caption{Summary of our probes, used prompts in each probe, purposes, and the number of groups using this probe for specific purpose.
}
\label{tab:prompts_usage}
\renewcommand{\arraystretch}{1.3}
\scalebox{0.9}{
\begin{tabular}{p{3cm}p{3cm}p{7cm}>{\centering\arraybackslash}p{2cm}}
\toprule
\textbf{Probe}                        & \textbf{Prompts}                                                      & \textbf{Purpose}                                                                                                               & \textbf{Group Usage} \\ \midrule
\textbf{Image Generation}             & Click a button and say: \textit{``Please give me a picture of ...''}                                       & \textbullet \, Generated images of attractions and food to fill out the timetable in Task 1.                                                     & 12              \\
                                     &                                                                        & \textbullet \, Generated images of attractions unfamiliar to the other generation to negotiate travel plans in Task 1.                                                  & 3               \\
                                     &                                                                        & \textbullet \, Generated images of campus facilities to better share their campus life in Task 2.                                                                             & 7               \\
                                     &                                                                        & \textbullet \, Generated images of working places to describe their current jobs and future aspirations.                                        & 2               \\ \midrule
\textbf{3D Objects Generation}        & Click a button and say: \textit{``Please give me ... (e.g., ``Bicycle'', ``Pencil box'', ``Chalk'', etc...) ''}                                                       & \textbullet \, Generated 3D models of vehicles to fill out the timetable in Task 1.                                                               & 12              \\
                                     &                                                                        & \textbullet \, Generated 3D models of stationery items to share campus life in Task 2.                                                           & 7              \\ \midrule
\textbf{Scene Generation}             & Click a button and say: \textit{``Please take me to ... (e.g., ``farm'', ``old school'', ``old classroom'', etc...) ''}                                                  & \textbullet \, Engaged in farm or old classrooms scene in VR to reminiscence and assist storytelling.                                                                       & 10              \\
                                     &                                                                        & \textbullet \, Switched from new and old classrooms in VR to discuss and compare their school life.                                                           & 6              \\ \midrule
\textbf{Info Assistant}               & Click a button and verbally describe the needs                                & \textbullet \, Asked for AI suggestion of travel plan in Task 1.                                                                                                  & 9               \\
                                     &                                                                        & \textbullet \, Queried information that both parties are not familiar with (e.g., ``Where is this section of the Great Wall located?'').                                                               & 2               \\
                                     &                                                                        & \textbullet \, Sought authoritative and systematic explanations of specific words (e.g., ``nèi juǎn'').                                                                              & 2               \\ \midrule
\textbf{Chat Facilitator}                & [Automatically] provide intervention based on communication contexts to provide appropriate voice suggestions
& \textbullet \, Suggested topics based on contextual content, helping two generations avoid awkward silences         & 8               \\ &
& \textbullet \, Provided timely interventions to maintain a balanced conversation  & 6               \\ \midrule
\textbf{Emotional Visualization}      & [Automatically] activate when detecting keywords     & \textbullet \, Helped inter-generational participants in expressing their emotions more effectively.                                                         & 7               \\ \bottomrule

\end{tabular}}
\end{table}

\begin{figure}[h]
\centering
\includegraphics[width=0.9\linewidth]{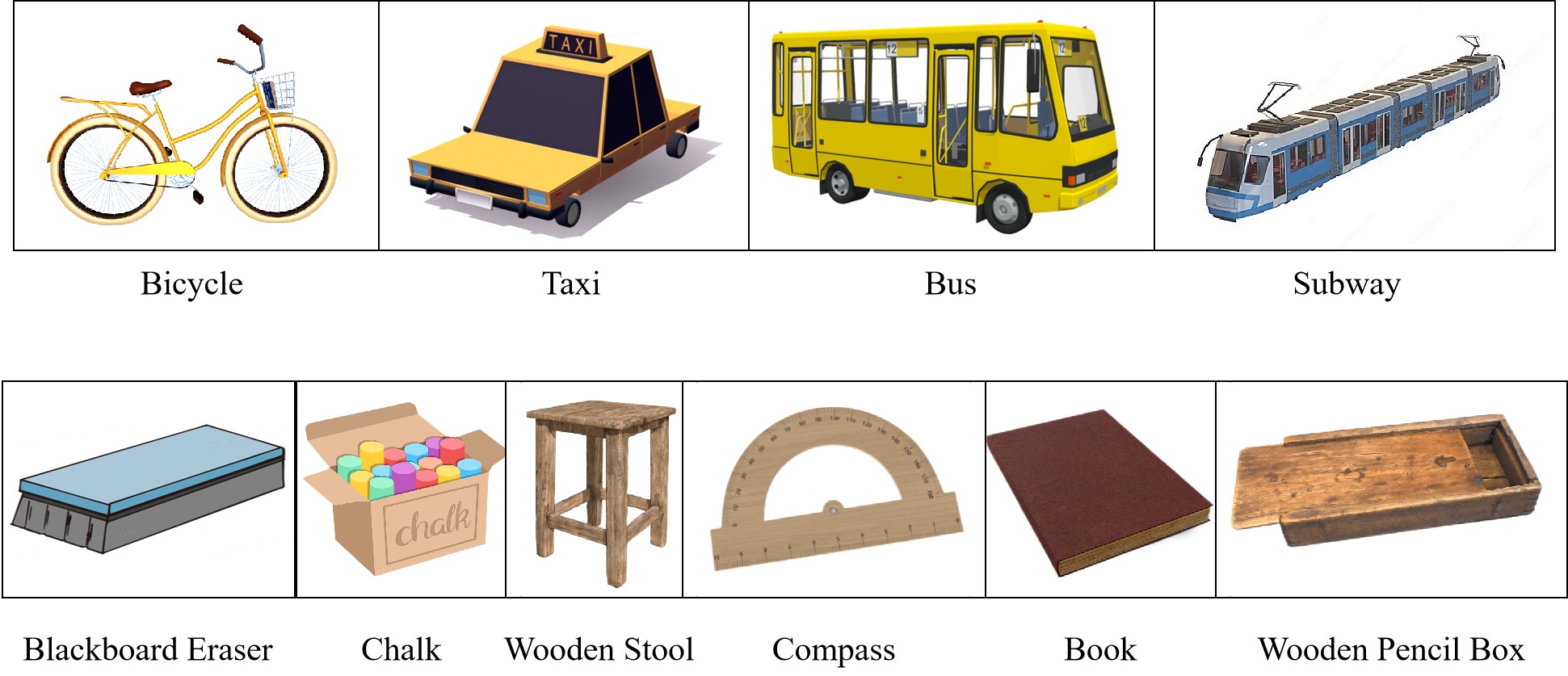}
\vspace{-0.2cm}
\caption{Pre-generated 3D objects} 
\label{fig:3dobjects}
\end{figure}

\begin{figure}[]
\centering
\includegraphics[width=0.8\linewidth]{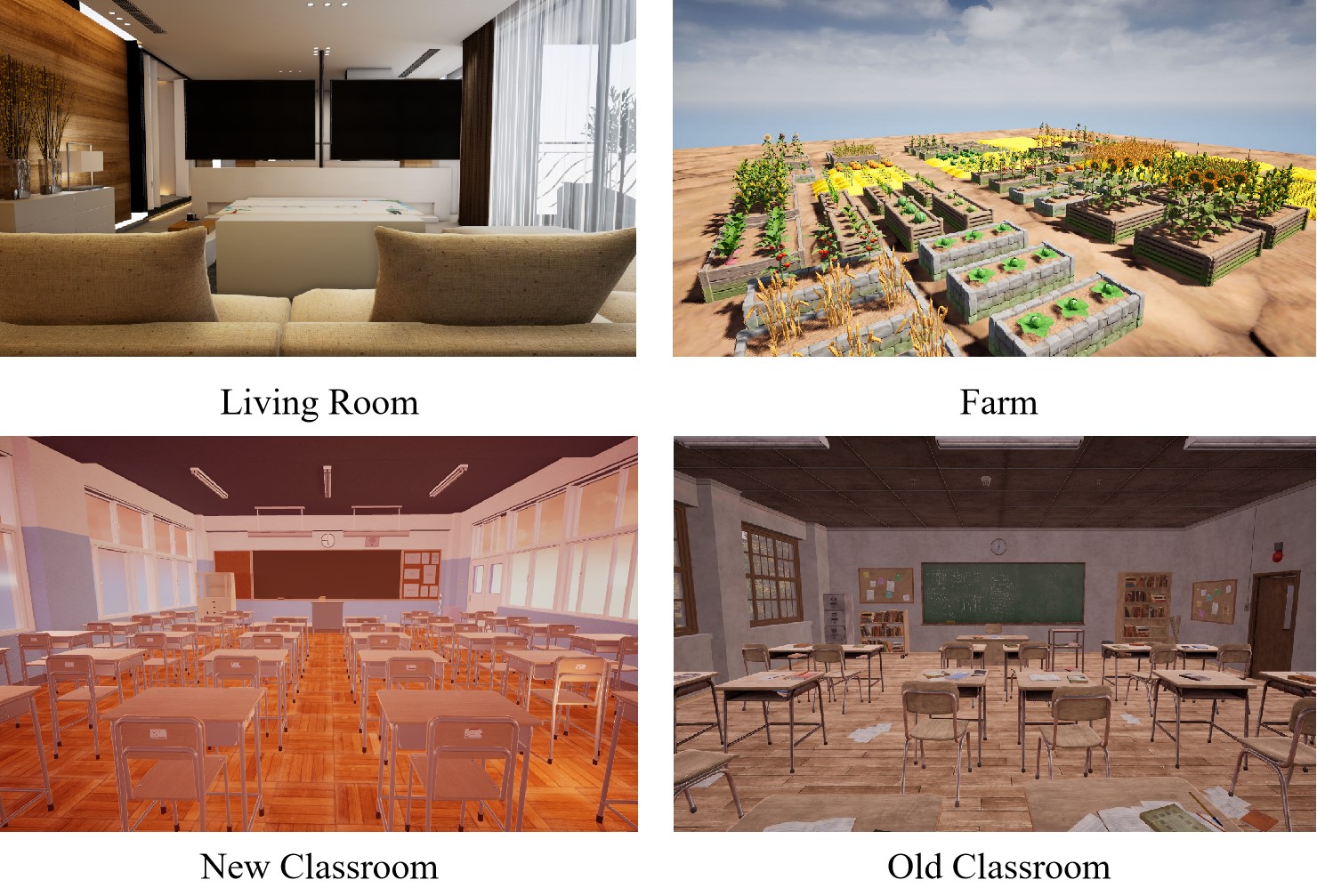}
\vspace{-0.2cm}
\caption{Pre-generated virtual scenes} 
\label{fig:scene}
\end{figure}

\begin{figure}[]
\centering
\includegraphics[width=0.8\linewidth]{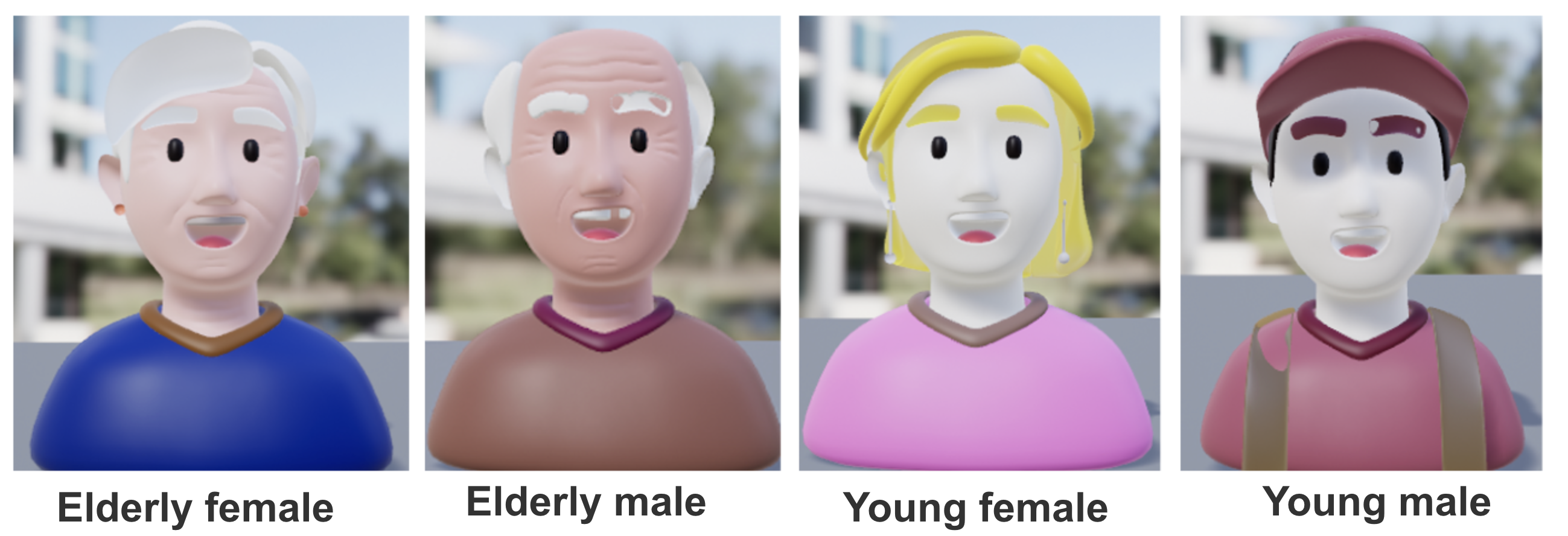}
\vspace{-0.2cm}
\caption{Four types of avatars} 
\label{fig:avatar}
\end{figure}

\end{document}